\documentclass[traditabstract]{aa}
\usepackage[varg]{txfonts}
\usepackage{graphics}

\def\cc{\,{\rm cm^{-3}}}
\def\cm2{\,{\rm cm^{-2}}}
\def\pc2{\,{\rm pc^{2}}}
\def\kms{\,{\rm {km\,s^{-1}}}}
\def\kkms{\,{\rm {K\,km s^{-1}}}}
\def\co{\,{\rm ^{12}CO}}
\def\thirco{\,{\rm ^{13}CO}}
\def\h2{\,{\rm H_{2}}}

\def\ci{\hbox{{\rm [C {\scriptsize I}]}}} 
\def\cii{\hbox{{\rm [C {\scriptsize II}]}}}
\def\pii{\hbox{{\rm [P {\scriptsize II}]}}}
\def\feii{\hbox{{\rm [Fe {\scriptsize II}]}}}
\def\nii{\hbox{{\rm [N {\scriptsize II}]}}}
\def\niii{\hbox{{\rm [N {\scriptsize III}]}}}
\def\neii{\hbox{{\rm [Ne {\scriptsize II}]}}}
\def\nev{\hbox{{\rm [Ne {\scriptsize V}]}}}
\def\oi{\hbox{{\rm [O {\scriptsize I}]}}}
\def\oiii{\hbox{{\rm [O {\scriptsize III}]}}}

\def\sivi{\hbox{{\rm [Si {\scriptsize VI}]}}}
\def\siii{\hbox{{\rm [S {\scriptsize III}]}}}
\def\oiv{\hbox{{\rm [O {\scriptsize IV}]}}}

\def\hii{\hbox{{\rm H {\scriptsize II}}}}

\def\go{{G_o}}

\def\c{{\rm C^{\circ}}}
\def\mm{{\rm $\mu$m}}
\def\c10{CI($^3P_1\rightarrow ^3P_0$)} 
 
\def\cb21{CI($^3P_2\rightarrow ^3P_1)$} 
\def\etal{\rm et\,al.\ }
\def\eg{e.g.,\ }
\def\ie{i.e.,\ }
\def\be{\begin{equation}}
\def\ee{\end{equation}}
\def\bea{\begin{eqnarray}}
\def\eea{\end{eqnarray}}
\def\aua{{\rm A\&A,} }

\def\aar{{\rm A\&AR,} }
\def\apj{{\rm ApJ,} }
\def\aj{{\rm AJ,} }
\def\apjs{{\rm ApJS,} }
\def\apjl{{\rm ApJL,} }

\def\mnras{{\rm MNRAS,} }

\begin{document}

\title{The outflow of gas from the Centaurus A circumnuclear disk}

\subtitle{Atomic spectral line maps from Herschel-PACS and APEX}

\author{F.P. Israel\inst{1}
 \and   R. G\"usten\inst{2}
 \and   R. Meijerink\inst{1}
 \and   M.A. Requena-Torres\inst{2,3} 
 \and   J. Stutzki\inst{4} 
}

 \offprints{F.P. Israel}

 \institute{Sterrewacht Leiden, Leiden University, P.O. Box 9513,
             2300 RA Leiden, The Netherlands 
  \and       Max-Planck-Institut f\"ur Radioastronomie, 
             Auf dem H\"ugel 69, 53121 Bonn, Germany
  \and       Space Telescope Science Institute, 3700 San Martin Drive, 
             Baltimore, 21218 MD, USA
  \and       I. Physikalisches Institut der Universit\"at zu K\"oln, Z\"ulpicher Strasse 77, D-50937 K\"oln, Germany  
 }

\authorrunning{F.P. Israel et al. }

\titlerunning{Cen~A circumnuclear disk and outflow}

\date{Received ????; accepted ????}
 
\abstract{The physical state of the gas in the central 500 pc of
  NGC~5128 (the radio galaxy Centaurus~A - Cen~A), was investigated
  using the fine-structure lines of carbon $\ci$, $\cii$; oxygen
  $\oi$, $\oiii$, and nitrogen $\nii$, $\niii$ as well as the
  $\co$(4-3) molecular line. The circumnuclear disk (CND) is traced by
  emission from dust and the neutral gas ($\ci$ and $\co$). A gas
  outflow with a line-of-sight velocity of 60 $\kms$ is evident in
  both lines. The $\ci$ emission from the CND is unusually strong with
  respect to that from CO. The center of the CND ($R<90$ pc) is bright
  in $\oi$, $\oiii$, and $\cii$; $\oi\lambda$63$\mu$m emission
  dominates that of $\cii$ even though it is absorbed with optical
  depths $\tau$=1.0-1.5.  The outflow is well-traced by the $\nii$ and
  $\niii$ lines and also seen in the $\cii$ and $\oiii$ lines that
  peak in the center.  Ionized gas densities are highest in the CND
  (about 100 $\cc$) and low everywhere else.  Neutral gas densities
  range from 4000 $\cc$ (outflow, extended thin disk ETD) to 20 000
  $\cc$ (CND).  The CND radiation field ($G_{o}\approx$4) is weak
  compared to the ETD starburst field ($G_{o}\approx$40). The outflow
  has a much stronger radiation field ($G_{o}=130$).  The total mass
  of all the CND gas is $9.1\pm0.9\,\times10^{7}$ M$_{\odot}$ but the
  mass of the outflowing gas is only $15\%-30\%$ of that. The outflow
  most likely originates from the shock-dominated CND cavity
  surrounding the central black hole.  With a factor of three
  uncertainty, the mass outflow rate is $\approx2$ M$_{\odot}$
  yr$^{-1}$, a thousand times higher than the accretion rate of the
  black hole. Without replenishment, the CND will be depleted in
  15-120 million years. However, the outflow velocity is well below
  the escape velocity.}  

\maketitle
 
\section{Introduction}

Giant elliptical galaxies, including the nearest example NGC~5128 ($D$
= 3.84 Mpc - Harris $\etal$ 2010), frequently contain deeply embedded
disks of dust and gas, remnants of smaller gas-rich galaxies that have
fallen in. The embedded disks are a transient phenomenon. Eventually,
the gas will be consumed by (a) accretion onto a central black hole,
(b) expulsion in the form of jets and flows emanating from the
nucleus, or (c) the formation of new stars. In fact, each of these
processes occurs widely in a variety of galaxies. In (ultra)-luminous
infra-red galaxies (LIRGs and ULIRGs), high-rate star formation is an
important if not dominant mechanism of circumnuclear gas consumption.
In galaxies with very active nuclei (AGNs), significant accretion onto
the central black hole occurs whether or not the galaxy is actively
forming stars. In radio galaxies (RGs) and many AGNs narrowly
collimated jets emanate directly from the nucleus and travel at very
high, sometimes relativistic, speeds. However, such jets carry little
mass by themselves. Much higher masses characterize the atomic and
molecular gas outflows from intense circumnuclear star-bursts, such as
those in the nearby galaxies NGC~253 and M~82 (Bolatto et al. 2013;
Leroy et al. 2015). These outflows travel much slower than nuclear
jets and are driven by massive stellar winds. More powerful outflows,
apparently not driven by stellar winds, have been identified in a
variety of active galaxies primarily by their absorption of the host
galaxy nuclear continuum. These include ULIRG/AGNs (such as Mrk~231,
Feruglio et al. 2015), Seyfert galaxies (such as IC~5063, Morganti et
al. 2015), and brightest cluster galaxies (such as those in Abell
A1664 and A1835, Russell et al. 2014; McNamara et al. 2014). Very high
outflow velocities up to $1000\,\kms$ and molecular mass outflow
rates of several hundred solar masses per year have been claimed (see
\eg Cicone $\etal$ 2014, McNamara $\etal$ 2014, Feruglio $\etal$
2015, and references therein). In most of these extreme cases the
outflow is not spatially resolved but deduced from extended wings in
molecular line profiles, and the quoted outflow rate depends on the
essentially unknown CO-to-$\h2$ conversion factor. It could be
significantly lower, if the widely shared assumption that the CO
outflow is optically thick and similar to the dense molecular gas in
galaxy disks, is wrong. Indeed, Dasyra $\etal$ (2016) have shown that
the molecular outflow in IC~5063 is optically thin and much less
massive than earlier assumed by some of the same authors (Morganti
$\etal$ 2015).

Most of these galaxies are so distant that circumnuclear disks with
diameters less than a kiloparsec are hardly or not at all resolved,
and emission from outflows is likewise hard to detect, making it
difficult to derive physical conditions with a degree of confidence.
A rare exception is NGC~5128, host of the huge FR~I radio source
Centaurus A (Cen~A - see a review by Israel 1998). At a distance of
3.84 Mpc (Harris et al. 2010), it is more than an order of magnitude
closer than almost all other active galaxies and even four times
closer than the iconic Seyfert galaxy NGC~1068.  Moreover, all three
mechanisms of central (molecular) gas disk consumption (outflow,
accretion, star formation) are potentially present in its center.

Optical images show a prominent dark band crossing the elliptical
galaxy that is in reality the projection of an embedded warped, thin
disk of gas and dust (Dufour et al. 1979, Nicholson \etal 1992)
extending over several kiloparsecs (hereafter called the `extended
thin disk', or ETD) with a mass of $1.5\times10^{9}$ M$_{\odot}$, two
per cent of the enclosed dynamical mass (cf Israel 1998). The ETD
appears to be the remnant interstellar medium (ISM) of a medium-sized
late-type galaxy that was absorbed by the giant elliptical at most a
few hundred million years ago (Graham, 1979; Struve \etal
2010). Optical and UV images reveal large numbers of young, luminous
blue stars. A recent estimate of the ETD star formation rate is
$\sim$1.6 M$_{\odot}$ yr$^{-1}$ (Wykes $\etal$ 2015).

At the center of NGC~5128/Cen~A is an accreting supermassive black
hole of $5\times10^{7}$ M$_{\odot}$ (Neumayer 2010). The black hole is
surrounded and obscured by a very compact (400 pc) circumnuclear disk
(CND; Israel \etal 1990).  This CND has been imaged in CO by Espada
\etal (2009), and its physical parameters have been explored by Israel
et al. (2014, hereafter Paper I) with the {\it Herschel} HIFI and
SPIRE instruments and ground-based (sub)millimeter telescopes. They
found a CND total gas mass of $8.4\times10^{7}$ M$_{\odot}$, partly
excited by X-rays or shock-induced turbulence. The two submillimeter
\ci\ lines are unusually strong in the center of NGC~5128, with
intensities exceeding those of the adjacent CO lines.  PDR models
consistent with the observed $\co$ and $\thirco$ fluxes predict only a
fraction of such intensity.

\nobreak The fine-structure line emission from neutral and ionized
carbon atoms, as well as other species such as oxygen and nitrogen, is
an important key to understanding the properties of the ISM in the
region they originate from, as it provides almost all of the gas
cooling.  For instance, warm and relatively tenuous gas is traced by
ionized carbon (\cii\,) and oxygen (\oiii\,), warm and dense gas is
traced by both neutral and ionized carbon (\ci\ and \cii\,) and
neutral oxygen (\oi\,), and cold and dense gas is primarily traced by
carbon monoxide (CO). The two neutral carbon \ci\, lines at 370 and
609 \mm\ can be measured from the ground with some difficulty.
Observation of the other fine-structure lines, from \oi\ at 63\mm\ to
\cii\ at 158\mm\ requires the use of airborne or space-borne
platforms.

\nobreak In this paper, we present observations of these
fine-structure lines and a molecular CO line covering the CND and part
of the ETD in NGC~5128 in order to continue our study of the central
region begun in Paper I, and further investigate the physical
conditions applying to the wider surroundings of its super-massive
black hole.  Parkin $\etal$ (2012, 2014) have published far-infrared
line and continuum studies of a much larger part of the ETD than
mapped by us. We have little to add to their results on the ETD, and
we accept their conclusion that the ETD physical characteristics are
similar to those of PDRs in spiral galaxy disks.

In contrast, the dynamics and the excitation of the dense CND gas are
expected to reflect the processes occurring in the vicinity of the
nuclear supermassive black hole, such as black hole accretion and jet
expulsion. In particular, with our CO, $\ci$, and $\cii$ line
measurements, we have sampled essentially all carbon in the galaxy's
center, allowing us to deduce total masses, as well as the mass
fractions associated with the various ISM phases in a more accurate
way than was possible before.

\begin{table}
\scriptsize
\caption[]{Log of {\it Herschel} observations}
\begin{center}
\begin{tabular}{ccccr}
\noalign{\smallskip}     
\hline\hline
\noalign{\smallskip}
Instru- & Transi-       & OBSID      & Date  & Integr. \\
ment    & tion          &            & Y-M-D & (sec)  \\
\noalign{\smallskip}     
\hline
\noalign{\smallskip}     
PACS  & 72-210          & 1342202588 & 2010-08-11 &  936 \\
PACS  & 55-72           & 1342203444 & 2010-08-24 & 5663 \\
HIFI  & $\ci$ $J$=1-0 C  & 1342201089 & 2010-07-21 &  353 \\
HIFI  & $\ci$ $J$=1-0 NW & 1342201092 & 2010-07-21 &  112 \\
HIFI  & $\ci$ $J$=1-0 SE & 1342201094 & 2010-07-21 &  112 \\
HIFI  & $\ci$ $J$=2-1 C  & 1342201712 & 2010-07-30 &   88 \\
HIFI  & $\ci$ $J$=2-1 NW & 1342201713 & 2010-07-30 &   88 \\
HIFI  & $\ci$ $J$=2-1 SE & 1342201716 & 2010-07-19 & 1884 \\
HIFI  & $\cii$ C         & 1342213717 & 2011-02-04 & 8970 \\
HIFI  & $\cii$ NW        & 1342201643 & 2010-07-28 &  936 \\
HIFI  & $\cii$ SE        & 1342201644 & 2010-07-28 &  936 \\
HIFI  & $\nii$           & 1342201778 & 2010-07-31 &  833 \\ 
SPIRE & SSPEC           & 1342204037 & 2010-08-23 & 5041 \\
\noalign{\smallskip}     
\hline
\noalign{\smallskip}
\end{tabular}
\end{center}
\label{herschellog}
\end{table}

\section{Observations and data handling}

\subsection{Herschel Space Observatory}

\nobreak All {\it Herschel} \footnote{{\it Herschel}
  is an ESA space observatory with science instruments provided by
  European-led Principal Investigator consortia and with important
  participation from NASA} (Pilbratt et al. 2010) observations with
the HIFI and PACS instruments described in this paper were obtained as
part of the Guaranteed Time Key Programme HEXGAL (PI:
R. G\"usten).  The {\it Herschel} observations using HIFI (Heterodyne
Instrument for the Far Infrared; de Graauw et al. 2010) and those
using SPIRE-FTS (Spectral and Photometric Imaging Receiver and
Fourier-Transform Spectrometer; Griffin \etal 2010) have been
described in our previous paper on the NGC~5128 CND (Israel \etal
2014).  A summary of all {\it Herschel} fine-structure line
observations, including those already presented by Israel \etal 2014,
is given in Table\,\ref{herschellog}.

\begin{figure}[]
\unitlength1cm
\begin{minipage}[t]{9cm}
\resizebox{9cm}{!}{\rotatebox{0}{\includegraphics*{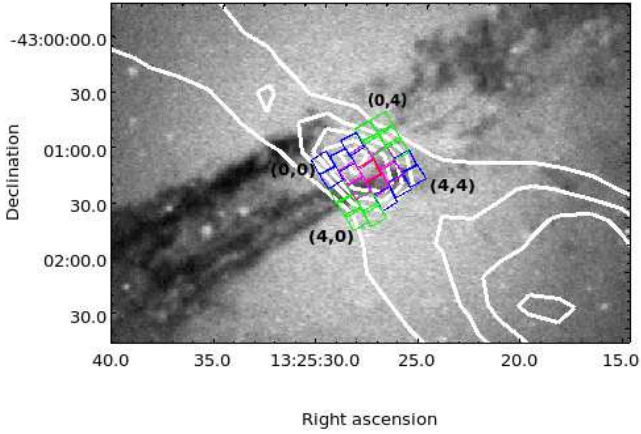}}}
\end{minipage}
\caption[]{The {\it Herschel} PACS footprint (colored squares) overlay
  on the SDSS-gray-tone image of the NGC~5128 central region. The
  different colors have no specific meaning. White contours trace the
  21cm radio continuum emission from the Centaurus A nucleus and
  jets. The array is oriented along the dark band image of
  the extended think disk. The circumnuclear disk is not
  distinguishable in this image, but is oriented at right angles to
  the jet direction. In the image, $1'$ corresponds to 1.1 kpc.}
\label{overlay}
\end{figure}

We have used the Photo-detector Array Camera $\&$ Spectrometer (PACS,
Poglitsch \etal 2010) on-board {\it Herschel} to map the distribution
of emission from various far-infrared atomic fine-structure lines in
NGC~5128. The PACS array consisted of 5$\times$5 `spaxels', each
of $9.4"$ size, combining into a field of $47"\times47"$ which
corresponds to 872$\times$872 pc at the distance of NGC~5128.  The
resolution of the {\it Herschel} PACS array varied from $9.5''$ to
$13"$ for wavelengths increasing from 60$\mu$m to 180$\mu$m, and was
thus undersampled by the individual spaxels.  Both PACS
observations were carried out in stare mode using chopping and
nodding. The array was pointed at the nominal position right ascension
(J2000) = 201.3650, declination (J2000) = -43.0191, with a pointing
accuracy of $2''$ r.m.s.. It was rotated over an angle of
$308.9^{\circ}$, thus orienting the array columns in a position angle
$129^{\circ}$ (more or less) parallel to the extended thin disk (ETD)
axis (see Figure\,\ref{overlay}, but we note that the major axis of the
CND is at a greater position angle ($P.A.\,=\,145^{\circ}$). We used range
spectroscopy to cover the full PACS wavelength range in two
observations. The first observation covered the wavelength ranges
70-105$\mu$m and 140-220$\mu$m, and the second covered the wavelength
ranges 51-73$\mu$m and 102-146$\mu$m (see Table\,\ref{herschellog}).
According to the on-line {\it Herschel} PACS observing manual, the
photometric calibration accuracy was about 11-12$\%$ r.m.s. The
absolute wavelength calibration was quoted as better than 20$\%$,
approaching 10$\%$ in band centers. However, for point-like sources,
the accuracy of the absolute wavelength was also determined by how well
the point source was centered on the corresponding spaxel; the 
wavelength-dependent error easily amounted to several tens of
kilometers per second.

\begin{figure}[]
\unitlength1cm
\begin{minipage}[t]{9cm}
\resizebox{9cm}{!}{\rotatebox{0}{\includegraphics*{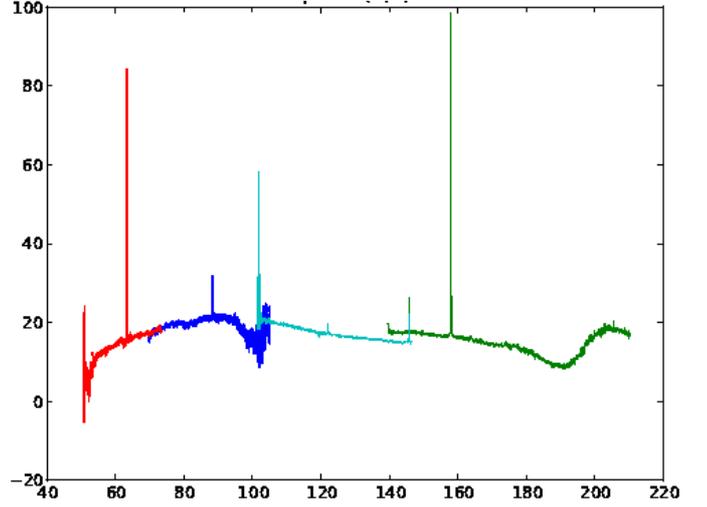}}}
\end{minipage}
\caption[]{Spectrum observed in the 175 pc sized central (2,2) spaxel
  of the PACS array, extending over the full range of wavelengths
  observed.  The emission lines from Table\,\ref{targets} are easily
  identified, as is the OH absorption line at a wavelength of 119
  $\mu$m. The four PACS spectral bands are indicated by different
  colors}
\label{spectrum}
\end{figure}

\begin{table}
\scriptsize
\caption[]{Spectral lines observed with {\it Herschel} PACS}
\begin{center}
\begin{tabular}{lrrcrr}
\noalign{\smallskip}     
\hline\hline
\noalign{\smallskip}
Line&Wavelength&Ionisation   & Beam$^{b}$ & \multicolumn{2}{c}{Spectral Resolution$^{b}$}\\
      &        & Potentials$^{a}$ & FWHM  & Wavelength  & Velocity \\
      &  ($\mu$m)  & (eV)        & ($"$) & ($\mu$m)    & ($\kms$) \\
\noalign{\smallskip}     
\hline
\noalign{\smallskip}     
\niii &  57.317    & 29.60-47.45 & 9.5  & 0.021  &  108 \\
\oi   &  63.184    & ---   13.62 & 9.5  & 0.017  &   88 \\
\oiii &  88.356    & 35.12-54.94 & 9.5  & 0.034  &  124 \\
\nii  & 121.898    & 14.53-29.60 & 10   & 0.116  &  300 \\
\oi   & 145.525    & ---   13.62 & 11   & 0.123  &  255 \\
\cii  & 157.741    & 11.26-24.38 & 11.5 & 0.126  &  240 \\
\nii  & 205.178    & 14.53-29.60 & 16   & 0.102  &  150 \\
\noalign{\smallskip}     
\hline
\noalign{\smallskip}
\end{tabular}
\end{center}
Notes: (a) Potentials required for creation and ionisation of the species, 
respectively (b) Taken from on-line PACS Observer's Manual, version 2.4.
\label{targets}
\end{table}

\begin{figure*}[]
\unitlength1cm
\begin{minipage}[t]{18cm}
\resizebox{6.95cm}{!}{\rotatebox{270}{\includegraphics*{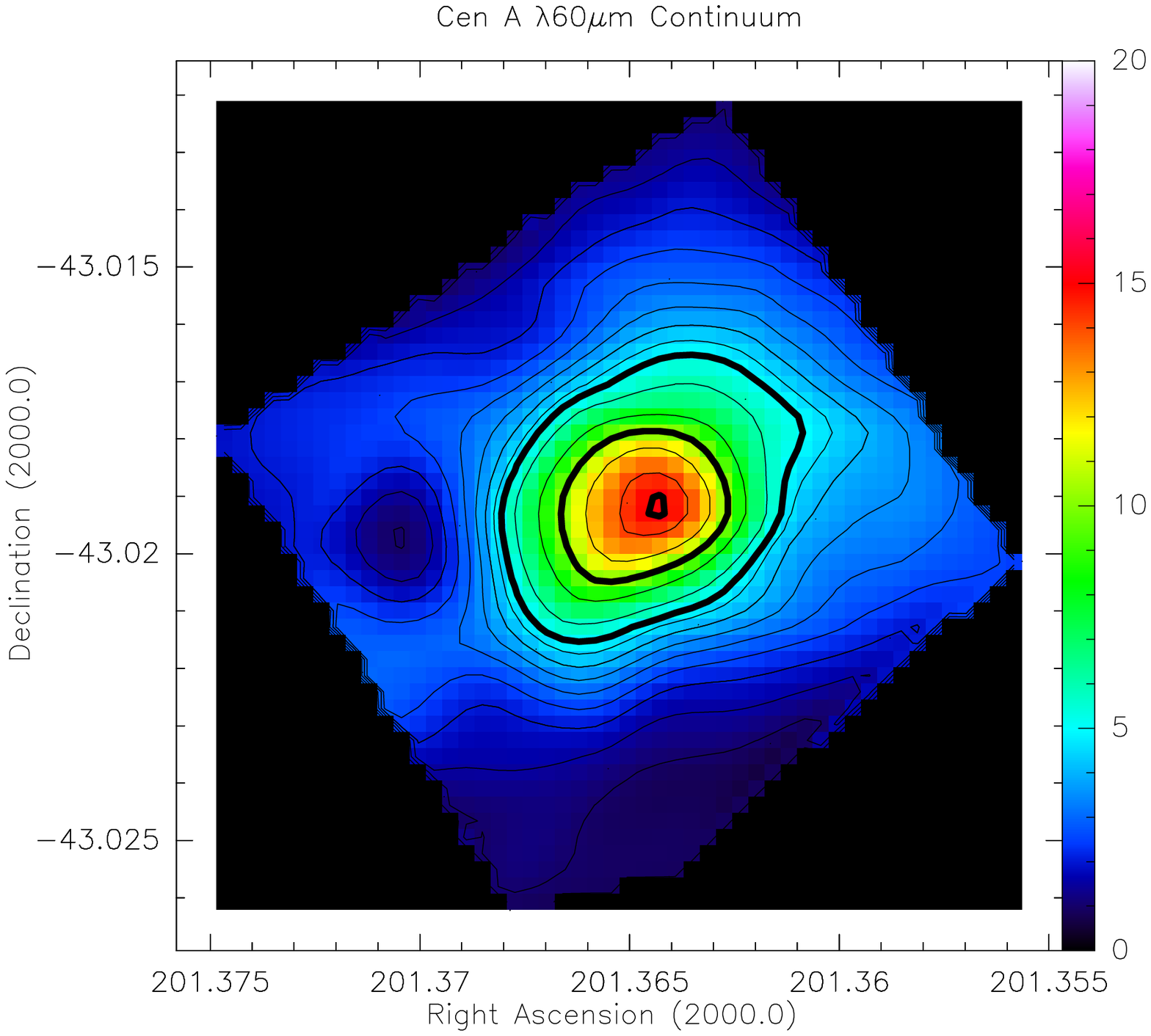}}}
\resizebox{5.6cm}{!}{\rotatebox{270}{\includegraphics*{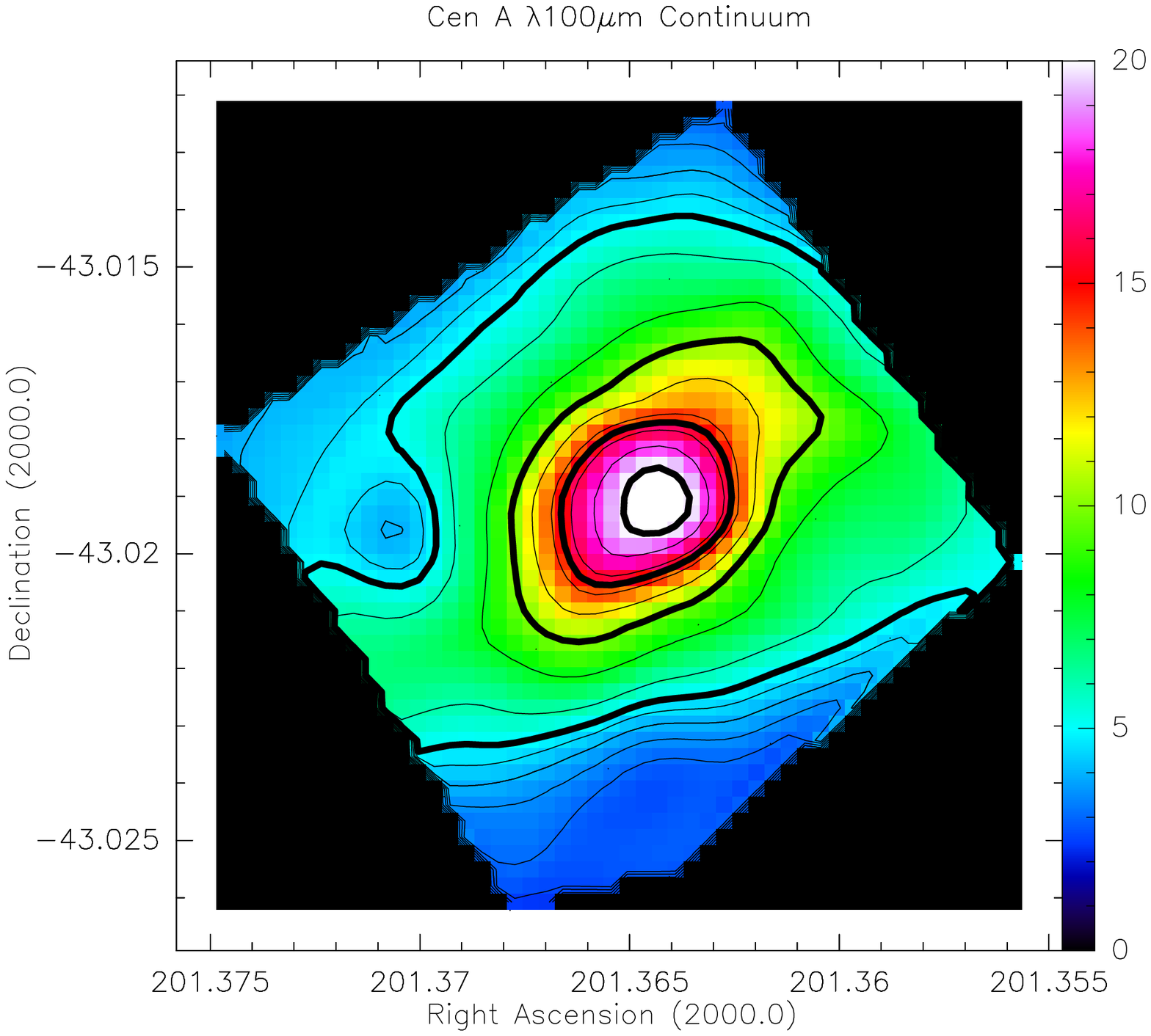}}}
\resizebox{5.6cm}{!}{\rotatebox{270}{\includegraphics*{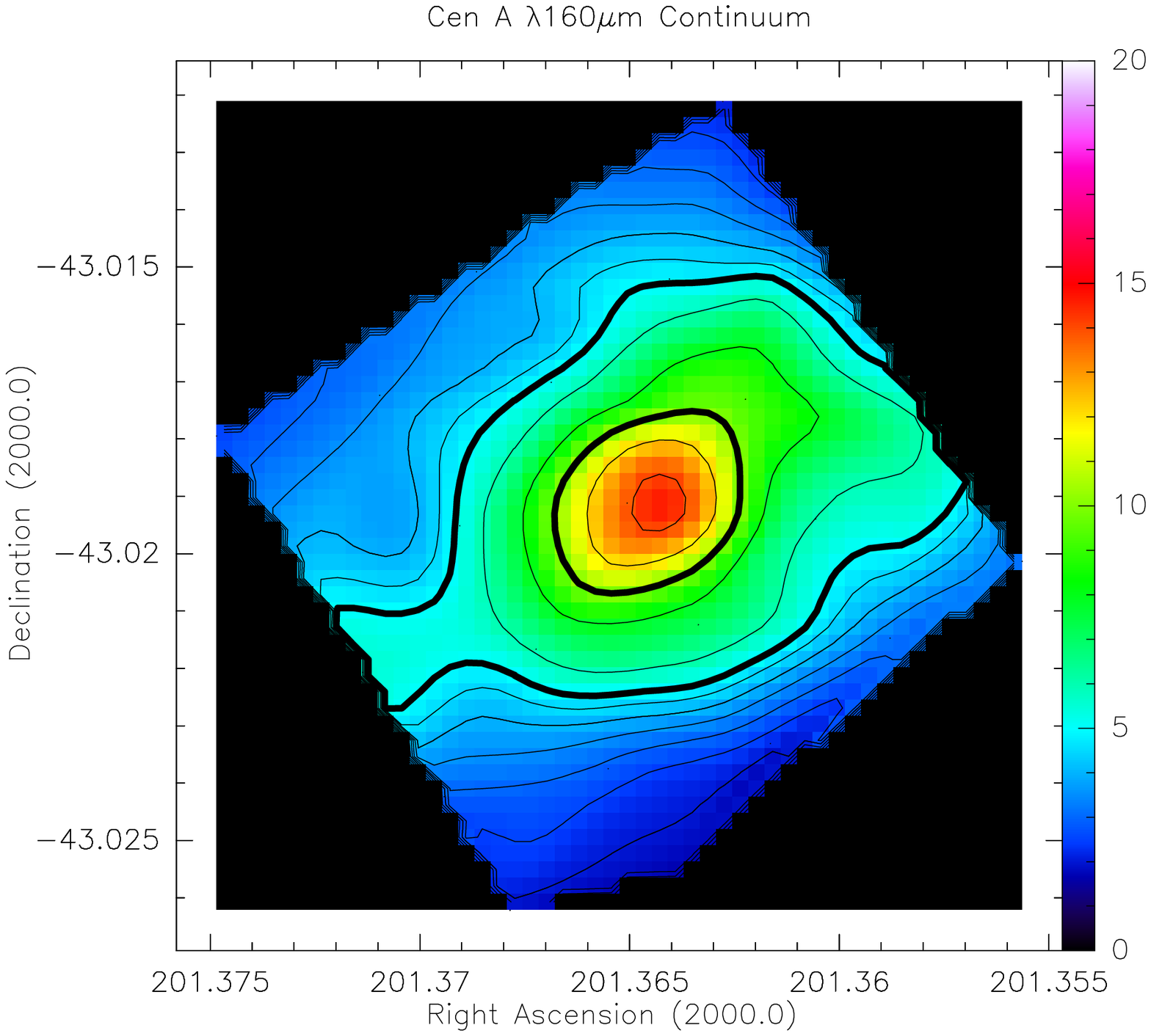}}}
\end{minipage}
\caption[]{Maps of the far-infrared continuum emission from the
  NGC~5128 central region. Left: 60$\mu$m map; center 100$\mu$m;
  right: 160$\mu$m map. All maps have contours at intervals of 0.5 Jy
  per PACS spaxel over the range 0.5-5 Jy/spax, and 2 Jy/spax over the
  range 6-20 Jy/spax. Heavy contours are at 5, 10, and 15 Jy/spax. The
  maps cover an area of roughly $0.01^{\circ}\times0.01^{\circ}$,
  corresponding to 0.67k pc $\times$ 0.67 kpc}
\label{contmaps}
\end{figure*}

The initial data processing was done using HIPE version 6.0. We used
the standard PACS pipeline for chop/nod observations of extended
sources, and version 16 of the PACS calibration. We then used
custom-made IDL and Python scripts to further analyze the data.
First, we extracted individual data cubes from the portions of the
spectra (cf. Fig.\,\ref{spectrum}) containing line emission. The lines
are identified in Table\,\ref{targets}. We removed linear baselines
from these cubes, excluding those regions that contained lines or
instabilities from the fit. This was sufficient for all the lines
except for the $\niii$57$\mu$m and $\nii$205$\mu$m lines that are
close to band edges (see Fig.\,\ref{spectrum}) where we had to
subtract a third-order baseline.  In addition, the absolute PACS
calibration of the $\nii$205$\mu$m was very uncertain because of filter
leakage and rapidly declining detector response (A. Poglitsch, private
communication) - see Fig.\,\ref{spectrum}.  We have therefore also scaled
extracted line fluxes to fit the SPIRE-derived $\nii$205$\mu$m
line flux (Israel \etal 2014) in the corresponding aperture.  Finally,
we have noted a strong and broad spectral feature in the
$\nii$122$\mu$m line spectra, displaced by about 2000 $\kms$ from the
systemic velocity. According to information from the Herschel on-line
PACS manual, this was a spectral ghost from the strong $\cii$158$\mu$m
line caused by a second pass in the optics of the PACS spectrometer.

\subsection{APEX 12m}

\nobreak We have used the Vertex Antennentechnik ALMA prototype
Atacama Pathfinder Experiment (APEX)\footnote{The Atacama Pathfinder
  Experiment (APEX) is a collaboration between the Max-Planck-Institut
  f\"ur Radioastronomie (MPIfR), the European Southern Observatory
  (ESO), and the Onsala Space Observatory (OSO).}  12-m telescope
(G\"usten \etal 2006) to observe the nucleus of NGC~5128 in the two
submillimeter \ci\ transitions at 492 and 809 GHz, as well as the
$\co$ $J$=4-3 transition at 461 GHz. The location of APEX at the high
elevation of 5105 m rendered it very suitable to high-frequency
observations from the ground. The observations were mostly made with
the First Light APEX Submillimeter Heterodyne (FLASH) dual-frequency
receiver (Heyminck \etal 2006, Klein \etal 2014) and the Carbon
Heterodyne Array (CHAMP+) receiver (G\"usten \etal 2008; Kasemann
\etal 2006), both developed by the Max Planck Institut f\"ur
Radioastronomie in Bonn (Germany); additional 492 GHz observations
were obtained with the Swedish Heterodyne Facility Instrument (SHeFI)
APEX-3 receiver (Vassilev $\etal$ 2008).  Main-beam efficiencies were
0.60 and 0.43 at operating frequencies of 464 and 812 GHz,
respectively. At the same frequencies, the antenna temperature to flux
density conversion factors were 48 and 70 Jy/K, respectively. APEX
FWHM beamwidths are $13.5"$ at 461 GHz, $12.7"$ at 492 GHz, and $7.7"$
at 809 GHz, a range very similar to the resolution of the {\it
  Herschel} PACS data described in this paper.

\nobreak All observations were made under excellent weather conditions
with typical overall system temperatures of 7500 K for CHAMP+-II (SSB,
800 GHz), and 500-800 K for FLASH-I (DSB, 460 \& 490 GHz).
Calibration errors were estimated at 15 to 20$\%$. Observations were
conducted with Fast Fourier Transform Spectrometer (FFTS; Klein et
al. 2006) back-ends for all instruments, except CHAMP+, where only the
two central pixels were attached to the FFTS back-ends. Other CHAMP+
pixels were attached to the MPI Array Correlator System (MACS)
backends. FFTS backends were able to reach resolutions of 0.12 MHz
(0.045 $\kms$ at 800 GHz), while the MACS units were used at a
resolution of 1 MHz (0.36 $\kms$ at 800 GHz). For the CHAMP+ data,
pointing was accurate within $\sim$5$''$. All observations were taken
in position switching mode with reference positions in azimuth ranging
from 600$''$ to 3600$''$.

The nuclear position was observed at various times between 2007 and
2011. We mapped the CND in four observing runs between Fall 2013
and Fall 2014 in both the $\ci$ 492 GHz and the $\co$ $J$=4-3 461 GHz
transitions.  The maps were made in a rectangular grid with full
sampling (grid spacing $6.5"$) and a major axis position angle PA =
145$^{\circ}$.

\begin{table}
\scriptsize
\caption[]{CND continuum flux densities}
\begin{center}
\begin{tabular}{rlrcr}
\noalign{\smallskip}     
\hline
\noalign{\smallskip}
$\lambda$ & Instrument & \multicolumn{3}{c}{Flux densities} \\
          &            & Map      &   CND+Nucleus & Nucleus \\
($\mu$m)  &            & \multicolumn{3}{c}{(Jy)}\\
\noalign{\smallskip}     
\hline
\noalign{\smallskip}
 20$^{a}$ &Spitzer &   -- &  3.3     & 2.4 \\
 30$^{a}$ &Spitzer &   -- &  6.4     & 2.8 \\
 60      & PACS  &   70 & 29$\pm$3 & 3.6 \\
 80      & PACS  &  108 & 41$\pm$6 & 4.0 \\
100      & PACS  &  144 & 50$\pm$8 & 4.3 \\
120      & PACS  &  131 & 43$\pm$7 & 4.6 \\
140      & PACS  &  119 & 39$\pm$7 & 4.8 \\
160      & PACS  &  111 & 35$\pm$6 & 5.1 \\
200$^{b}$ & SPIRE-SSW &   -- & 29       & 5.5 \\
300$^{b}$ & SPIRE-SSW &   -- & 15       & 6.4 \\
350$^{b}$ & SPIRE-LSW &   -- & 14       & 6.7 \\
450$^{b}$ & SPIRE-LSW &   -- & 10       & 7.4 \\
600$^{b}$ & SPIRE-LSW &   -- &  8.3     & 8.2 \\
\noalign{\smallskip}     
\hline
\noalign{\smallskip}
\end{tabular}
\end{center}
Notes: (a) From Weedman \etal (2005); (b)  From Israel \etal (2014)
\label{fluxes}
\end{table}

\begin{figure}[h]
\unitlength1cm
\begin{minipage}[t]{9cm}
\begin{center}
\resizebox{8cm}{!}{\rotatebox{0}{\includegraphics*{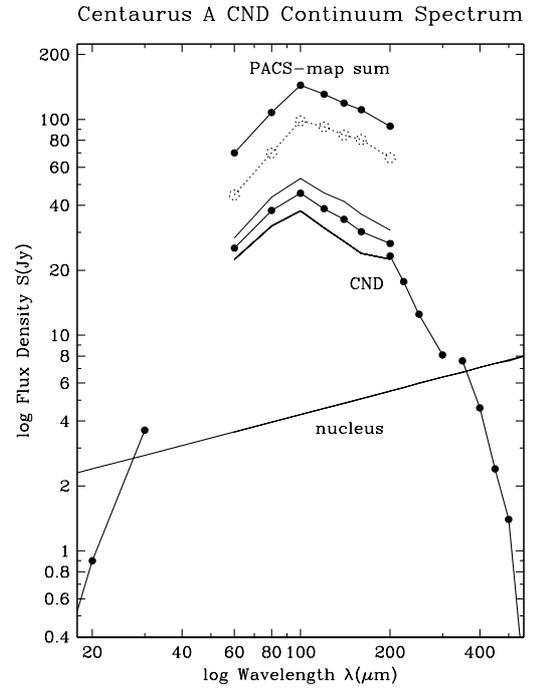}}}
\end{center}
\end{minipage}
\caption[]{Far-infrared spectra of the central disk in NGC~5128.  The
  curve at the top is the sum of all flux densities in the maps. The
  straight line is the (extrapolated) emission from the
  milli-arcsecond nuclear point source. The spectrum of the CND is
  corrected for the contribution of this nuclear point source; the four
  unconnected segments mark data from {\it Spitzer} LL, PACS,
  SPIRE-SWS, and SPIRE-LWS, respectively. Thin lines around the PACS
  segment indicate the uncertainty mostly caused by the difficulty of
  separating the peak from the extended surroundings.  The dotted line
  marks the spectrum of the emission in the map after subtraction of 
  both the CND and nucleus flex contributions.  }
\label{contspec}
\end{figure}

\section{Results }

\subsection{The CND far-infrared continuum spectrum}

The spatial resolution of the {\it Herschel} PACS observations is
sufficient to disentangle the circumnuclear disk far-infrared
continuum emission from both the strong emission from the extended
dust structures and from the nucleus itself.  Except for small
differences in resolution (see Table\,\ref{targets}), the continuum
maps shown in Fig.\,\ref{contmaps} resemble one another and also the
CO(2-1) map published by Espada \etal (2009).  At the PACS
resolution, there is continuum emission in each map spaxel. The lowest
levels in the map vary from 1 Jy/spax to 3 Jy/spax, typically about
$15\%$ of the peak emission in the map. Superposed on this base level
is an extended emission structure roughly twice as bright that
represents the so-called parallelogram structure that is seen in all
far-infrared and submillimeter maps. Both the base level emission and
the parallelogram emission are part of the ETD rather than the CND.

In the map centers, the bright and just-resolved emission from the CND
is an elliptical source in position angle $P.A.\,\approx\,45^{\circ}$,
where the position angle is counted counter-clockwise from north.  The
peak at the nuclear position is partly due to the non-thermal nuclear
point source coincident with the supermassive black hole.  We may
extrapolate the submillimeter nuclear flux densities summarized in
Section 3.2 of paper I with the power-law spectrum
$F_{\nu}\,\propto\,\nu^{-0.36}$ (Meisenheimer \etal 2007).  At the
wavelengths studied here, this milli-arcsecond nuclear source has a
flux density of about 4 Jy.  Taking into account this contribution, it
still appears that the dusty disk depicted in Fig.\,\ref{contmaps} is
significantly brighter in the center.  However, the characteristic
dimensions of the various continuum structures are close to the
resolution and pixel size of the PACS observations, and a more
accurate separation of the nuclear compact source, the slightly
extended CND, and the more extended ETD structures is hard to achieve.

In Fig.\,\ref{contspec} we show the spectrum of the continuum emission
summed over the entire PACS footprint, the extrapolated nuclear point
source spectrum, and the spectrum of the emission from the bright
circumnuclear disk only. In the CND spectrum we include the continuum
flux densities measured with {\it Herschel} SPIRE (Paper I) and {\it
  Spitzer} (Weedman \etal 2005). The CND flux densities are summarized
in Table\,\ref{fluxes}.  The SPIRE-SSW flux densities had a spatial
resolution of $18"$ and should provide a reasonably good
representation of the CND spectrum. The SPIRE-LSW flux densities were
obtained with apertures between $36"$ and $30"$ and were less accurate
as they will contain a varying contribution from the ETD in addition
to the CND flux. The {\it Spitzer} data were obtained with slits
$10"-20"$ wide, comparable to the CND dimensions.

\begin{table}
\scriptsize
\caption[]{APEX $J$=1-0 $\ci$ and $J$=4-3$\co$ map line fluxes}
\begin{center}
\begin{tabular}{rccc|rccc}
\noalign{\smallskip}     
\hline
\noalign{\smallskip}
Offsets & $I_{\ci}$ & $I_{CO}$ & Line & Offsets & $I_{\ci}$ & $I_{CO}$ & Line \\
$\Delta\alpha$,$\Delta\delta$ & fit      & fit       & ratio & $\Delta\alpha$,$\Delta\delta$ & fit      & fit       & ratio \\
($"$)& \multicolumn{2}{c}{($\kkms$)} &   & ($"$)& \multicolumn{2}{c}{($\kkms$)} &     \\
\noalign{\smallskip}     
\hline\hline
\noalign{\smallskip} 
-20.0 , +17.1 &  15.7 &   ... & ...  &   +5.3 ,  +3.7 &  86.2 &  59.0 & 1.46 \\ 
-16.2 , +11.8 &  39.7 &   ... & ...  &   +6.1 , -20.1 &   3.0 &   ... & ...  \\ 
-17.8 ,  +2.6 &  31.3 &   ... & ...  &   +7.2 , +12.3 &  32.2 &  21.3 & 1.5: \\ 
-14.9 , +21.5 &  15.5 &  30.5 & 0.51 &   +7.5 , -10.6 &  35.2 &  35.7 & 0.99 \\ 
-14.1 ,  -2.6 &  14.0 &  24.7 & 0.57 &   +8.9 , +21.3 &  28.7 &   ... & ...  \\ 
-12.8 ,  +6.9 &  14.8 &  40.5 & 0.37 &   +9.1 ,  -1.6 &  67.7 &  37.2 & 1.82 \\ 
-11.2 , +16.0 &   9.0 &   ... &  ... &   +9.7 , -25.5 &  18.5 &   ... & ...  \\ 
-10.6 ,  -7.5 &  11.0 &  23.7 & 0.47 &  +10.6 ,   7.5 &  30.3 &  31.0 & 0.98 \\ 
 -9.3 , +24.6 &  12.5 &   ... &  ... &  +11.2 , -16.0 &  29.2 &   ... & ...  \\ 
 -9.1 ,  +1.6 &  26.8 &  32.2 & 0.83 &  +12.5 , +16.1 &  20.8 &   ... & ...  \\ 
 -7.5 , +10.6 &  35.3 &  37.0 & 0.96 &  +12.8 ,  -6.9 &  26.2 &  31.0 & 0.84 \\ 
 -6.7 , -13.2 &   5.2 &  13.7 & 0.38 &  +14.6 ,  +1.7 &  20.5 &  29.5 & 0.69 \\ 
 -5.6 , +19.3 &  13.7 &   ... &  ... &  +14.9 , -21.3 &  10.0 &  18.7 & 0.54 \\ 
 -5.3 ,  -3.7 &  36.8 &  60.7 & 0.61 &  +16.2 , +10.8 &  12.7 &   7.8 & 1.8: \\ 
 -3.7 ,  +5.3 &  71.2 &  52.8 & 1.35 &  +16.8 , -12.7 &  19.2 &   ... & ...  \\ 
 -3.0 , -18.6 &   1.3 &   ... & ...  &  +18.4 ,  -3.6 &  11.7 &   ... & ...  \\ 
 -2.1 ,  14.4 &  25.7 &  36.0 & 0.71 &  +17.8 , +19.8 &  12.5 &   ... & ...  \\ 
 -1.6 ,  -9.1 &  37.5 &  38.3 & 0.98 &  +20.0 ,  +5.4 &  19.7 &  17.0 & 1.3: \\ 
    0 ,     0 & 109.0 &  75.5 & 1.43 &  +20.5 , -18.1 &  12.2 &   ... & ...  \\ 
 +1.6 ,  +9.1 &  51.5 &  45.7 & 1.13 &  +21.5 , +14.5 &  29.3 &   ... & ...  \\ 
 +2.1 , -14.4 &   9.0 &  35.2 & 0.27 &  +23.7 ,  +0.1 &  16.0 &   ... & ...  \\ 
 +3.5 , +17.7 &  22.3 &   ... & ...  &  +25.3 ,  +9.1 &   4.8 &   ... & ...  \\ 
 +3.7 ,  -5.3 &  64.0 &  52.5 & 1.22 &                &       &       &      \\ 
\noalign{\smallskip}     
\hline
\noalign{\smallskip}
\end{tabular}
\label{carbonmapdat}
\end{center}
\end{table}

\begin{figure*}[]
\unitlength1cm
\begin{minipage}[t]{18cm}
\resizebox{6.24cm}{!}{\rotatebox{270}{\includegraphics*{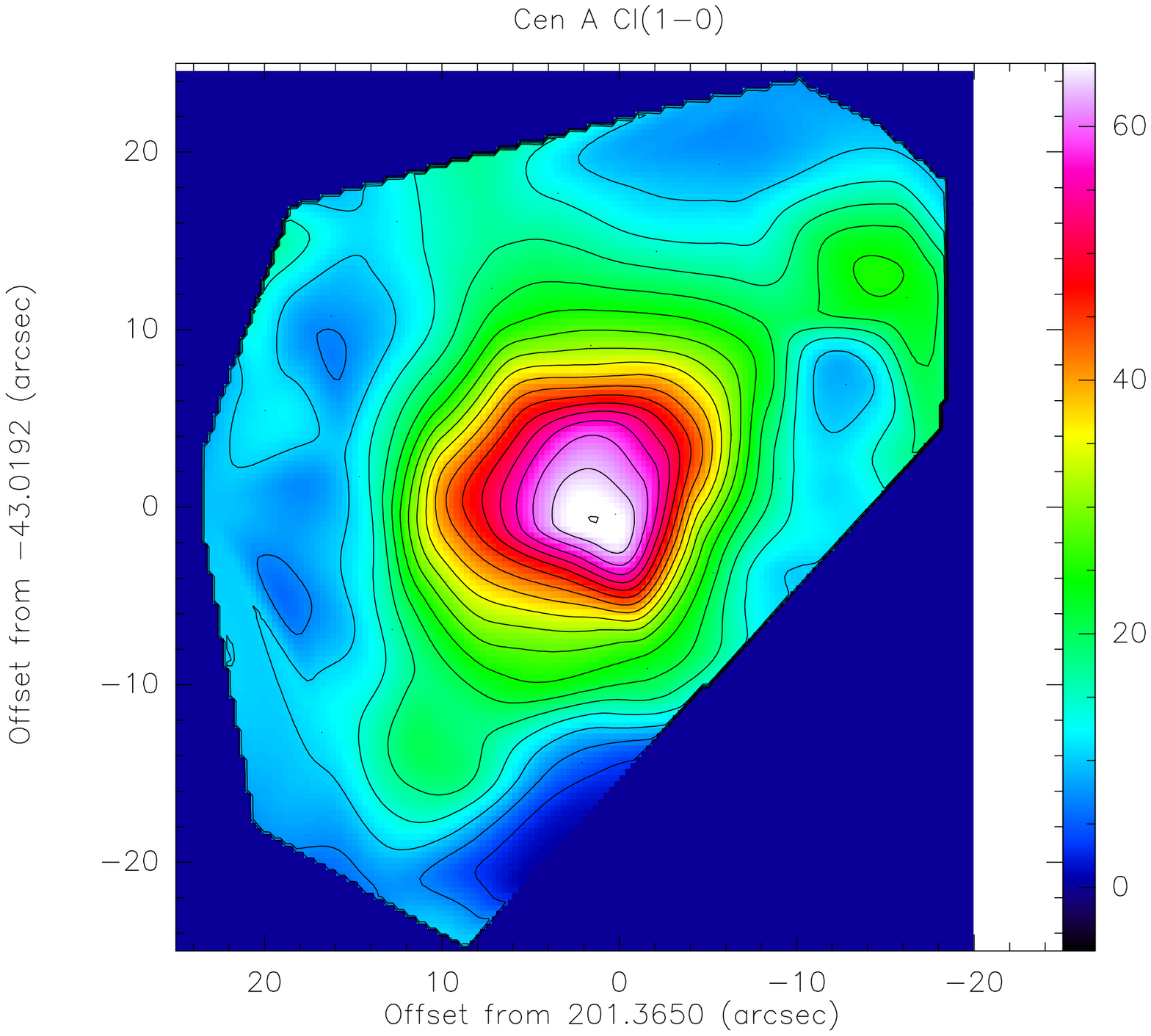}}}
\resizebox{6.24cm}{!}{\rotatebox{270}{\includegraphics*{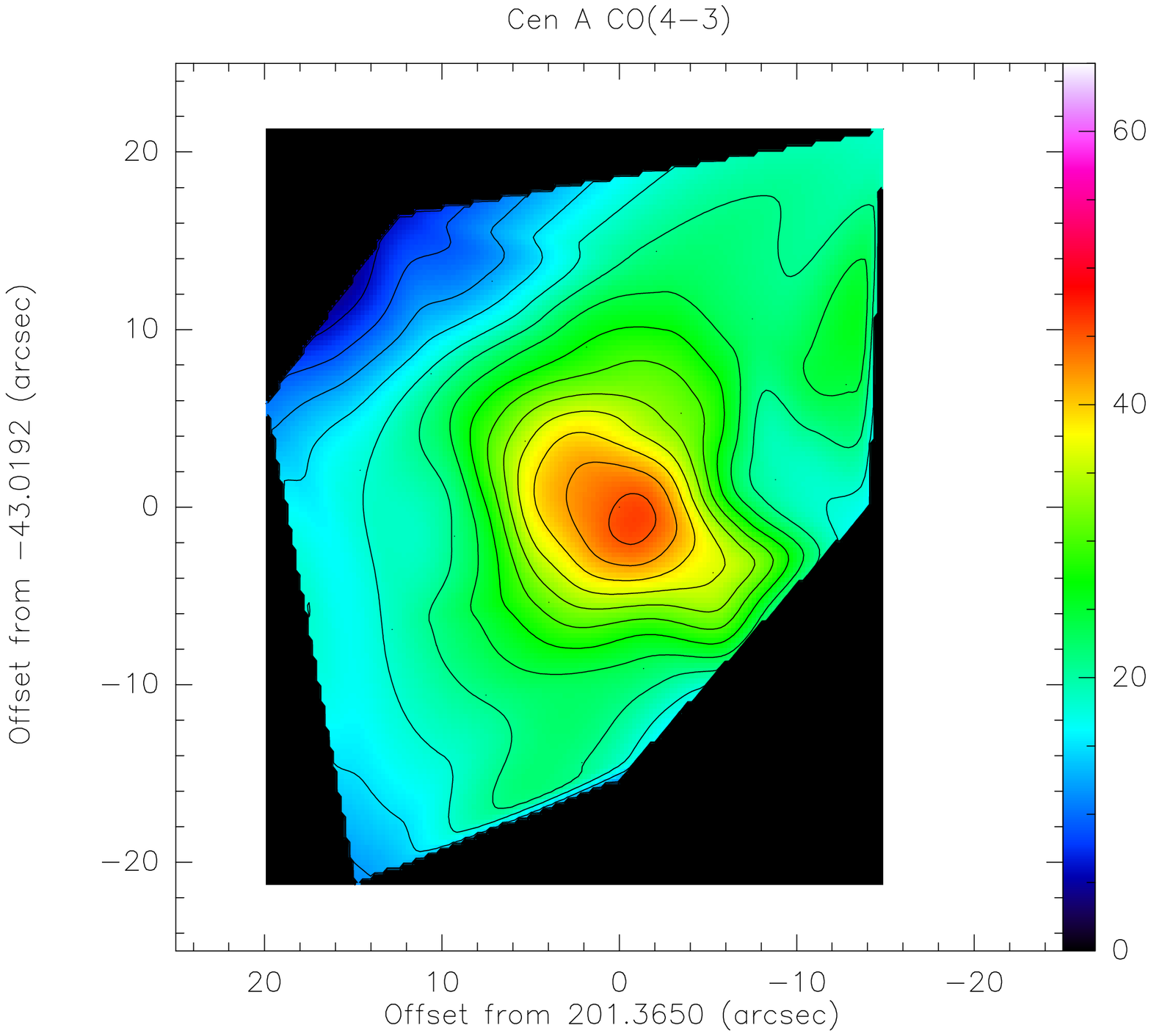}}}
\resizebox{6.24cm}{!}{\rotatebox{270}{\includegraphics*{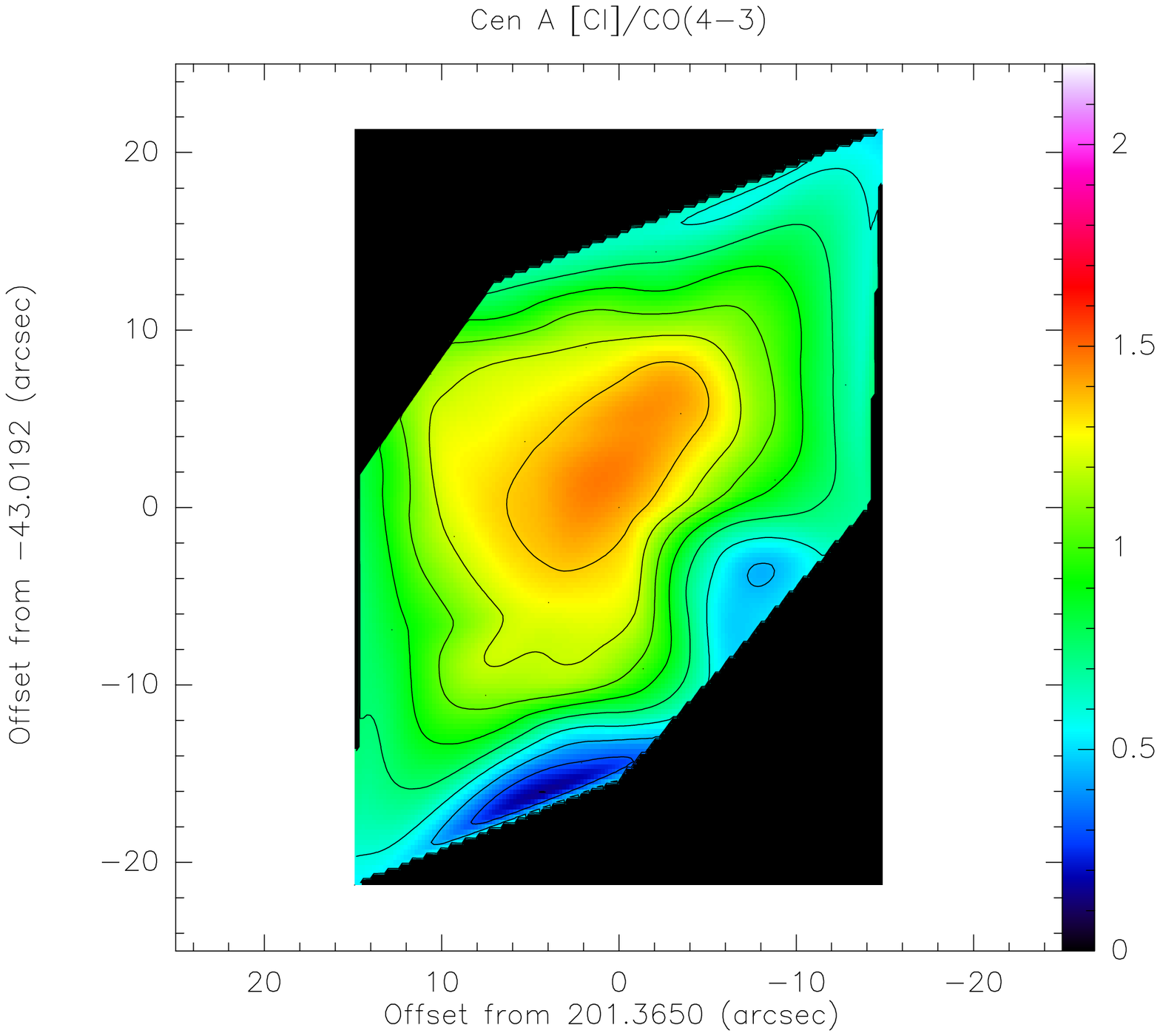}}}
\end{minipage}
\caption[]{Maps of the fitted $J$=4-3 $\co$ (left panel) and $J$=1-0
  [CI] line (middle panel) integrated main-beam brightness
  temperatures towards the center of Cen~A.  The fitted maps are
  almosat completely insensitive to the absorption against the
  continuum nucleus that would otherwise depress the line fluxes at
  the center.  Contour values are in steps of 6 $\kkms$ for [CI] and 5
  $\kkms$ for CO and peak at 110 $\kkms$ and 77 $\kkms$, respectively
  (the color wedge is in units of T$_{\rm A}^*$ = 0.6 T$_{\rm mb}$).
  The [CI]/CO ratio map (right panel) has contours at intervals of
  0.15 which clearly reveal the outline of the CND and its
  northeastern extension.}
\label{carbonmapfig}
\end{figure*}

\begin{figure*}[]
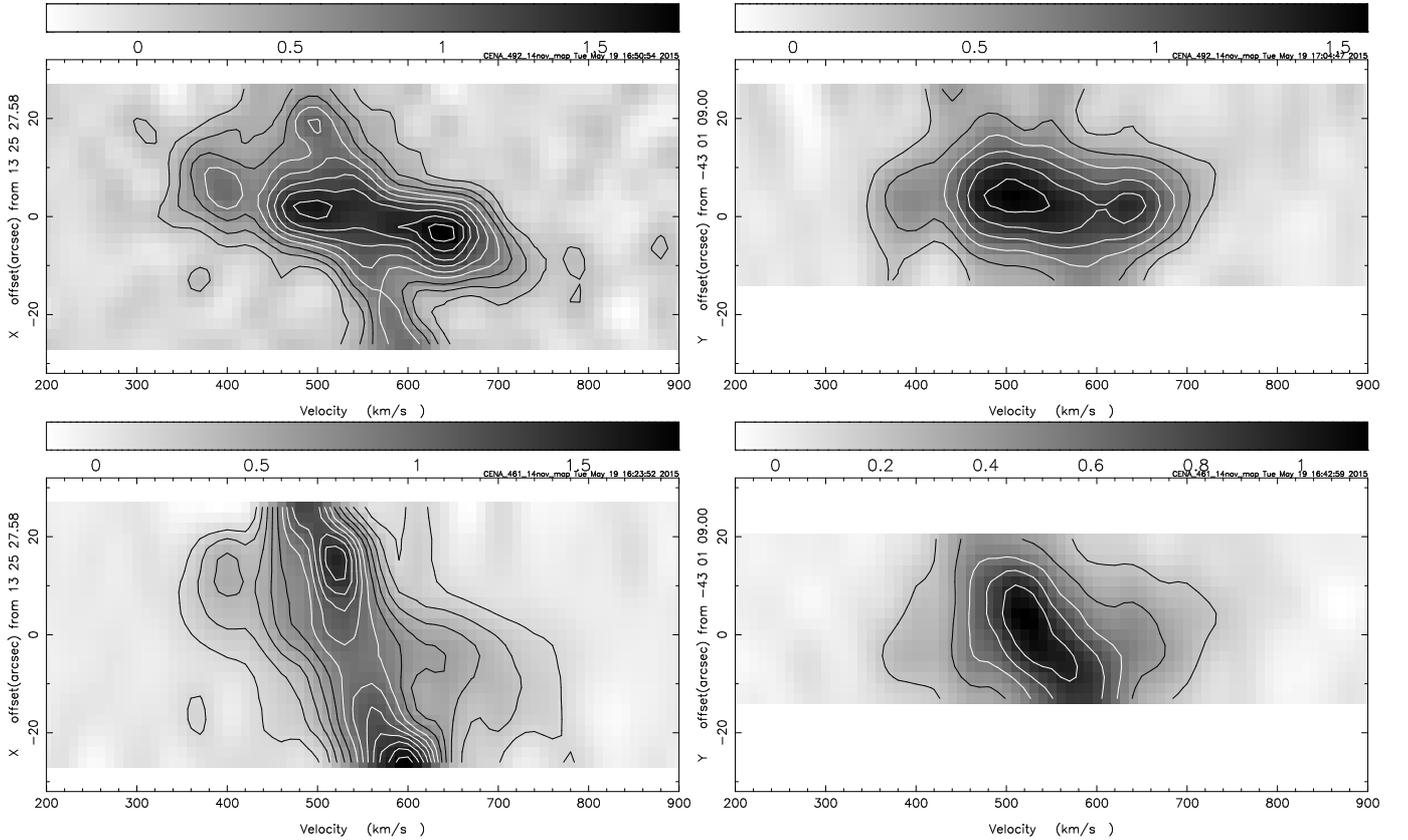

\unitlength1cm
\begin{minipage}[t]{18cm}
\resizebox{9cm}{!}{\rotatebox{270}{\includegraphics*{CENA_492_CND_40mK.eps}}}
\resizebox{9cm}{!}{\rotatebox{270}{\includegraphics*{CENA_492_OUT_40mK.eps}}}
\end{minipage}
\begin{minipage}[t]{18cm}
\resizebox{9cm}{!}{\rotatebox{270}{\includegraphics*{CENA_461_CND_30mK.eps}}}
\resizebox{9cm}{!}{\rotatebox{270}{\includegraphics*{CENA_461_OUT_40mK.eps}}}
\end{minipage}
\caption[]{Position-velocity diagrams of the observed emission in
  $J$=(1-0) $\ci$ (top panels) and $J$=4-3 $\co$ (bottom panels)
  Panels on the left show the velocity distribution along the CND
  major axis, panels on the right the velocity distribution along a
  line perpendicular to this, corresponding to the position angle of
  the radio/X-ray jet. The velocity scale is $V_{LSR}$.  Emission from
  the CND is at $X$ = $\pm10"$ and $V_{\rm LSR}$ = 550$\pm$200 $\kms$,
  emission from the ETD extends over the full $X$-range but is limited
  to $V_{\rm LSR}$ = $550\pm50$ $\kms$. The contours in the $J$=4-3 $\co$
  map at lower left are at multiples of 30 mK in $T_{mb}$, and at
  multiples of 40 mK in all other panels. The relatively low
  brightness at the center of all panels is due to the strong line
  absorption against the nuclear continuum. When this is taken into
  account, all diagrams show a central peak at $X,Y$ = 0,
  $V_{\rm LSR}$ = 550 $\kms$ instead.  }
\label{pvplots}
\end{figure*}

\begin{figure*}[]
\unitlength1cm
\begin{flushright}
\begin{minipage}[t]{18cm}
\resizebox{6.8cm}{!}{\rotatebox{270}{\includegraphics*{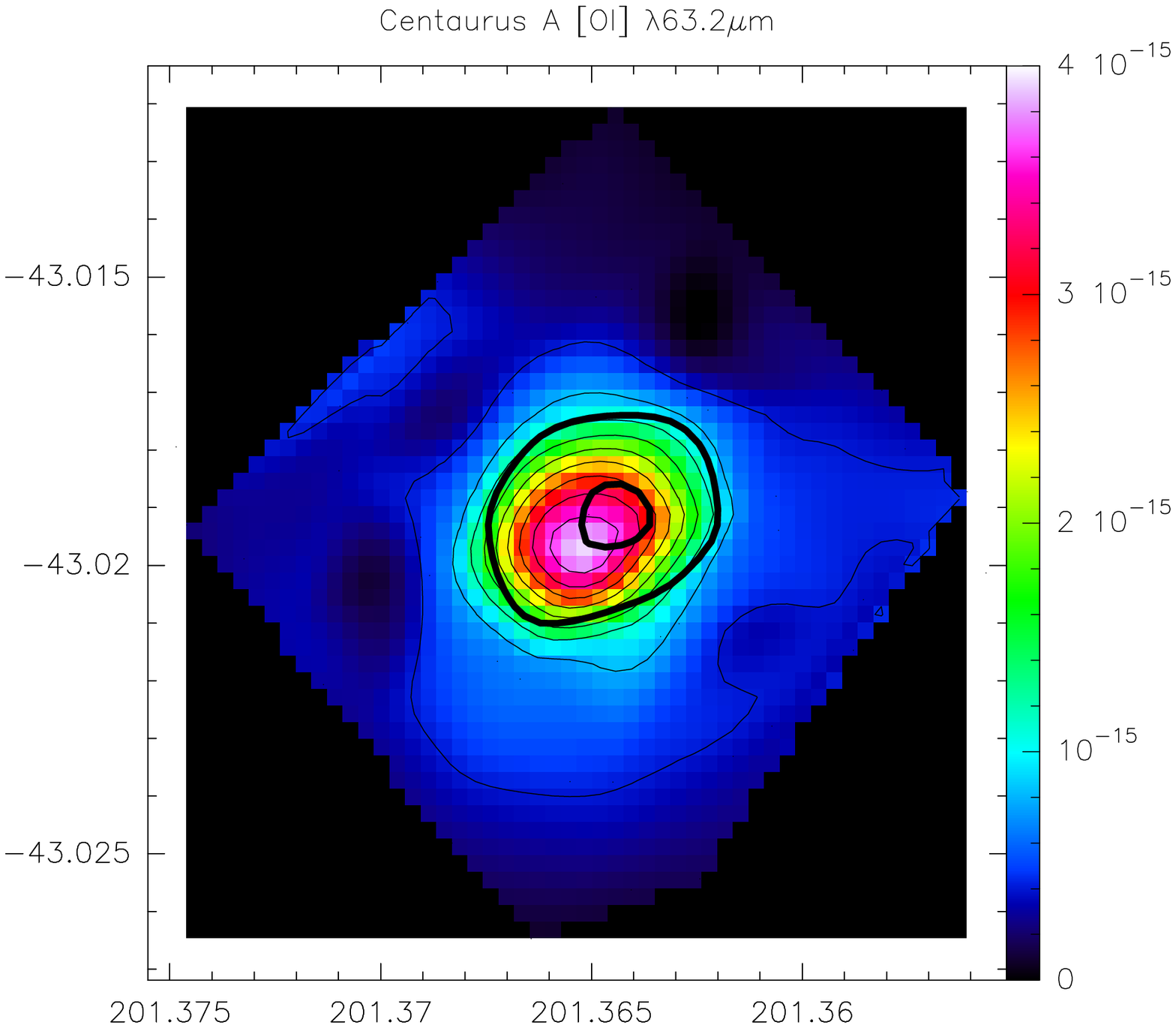}}}
\resizebox{5.55cm}{!}{\rotatebox{270}{\includegraphics*{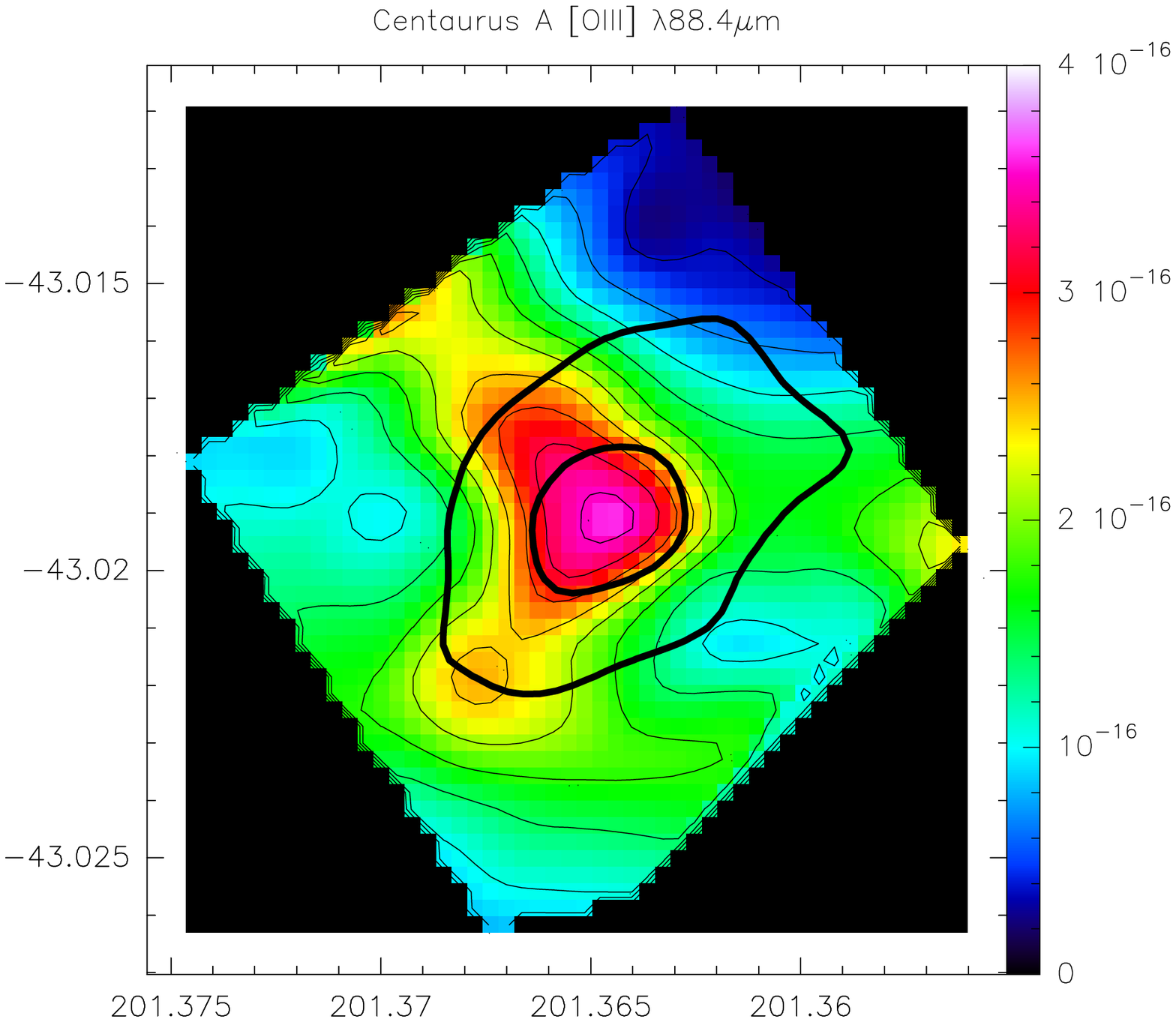}}}
\resizebox{5.8cm}{!}{\rotatebox{270}{\includegraphics*{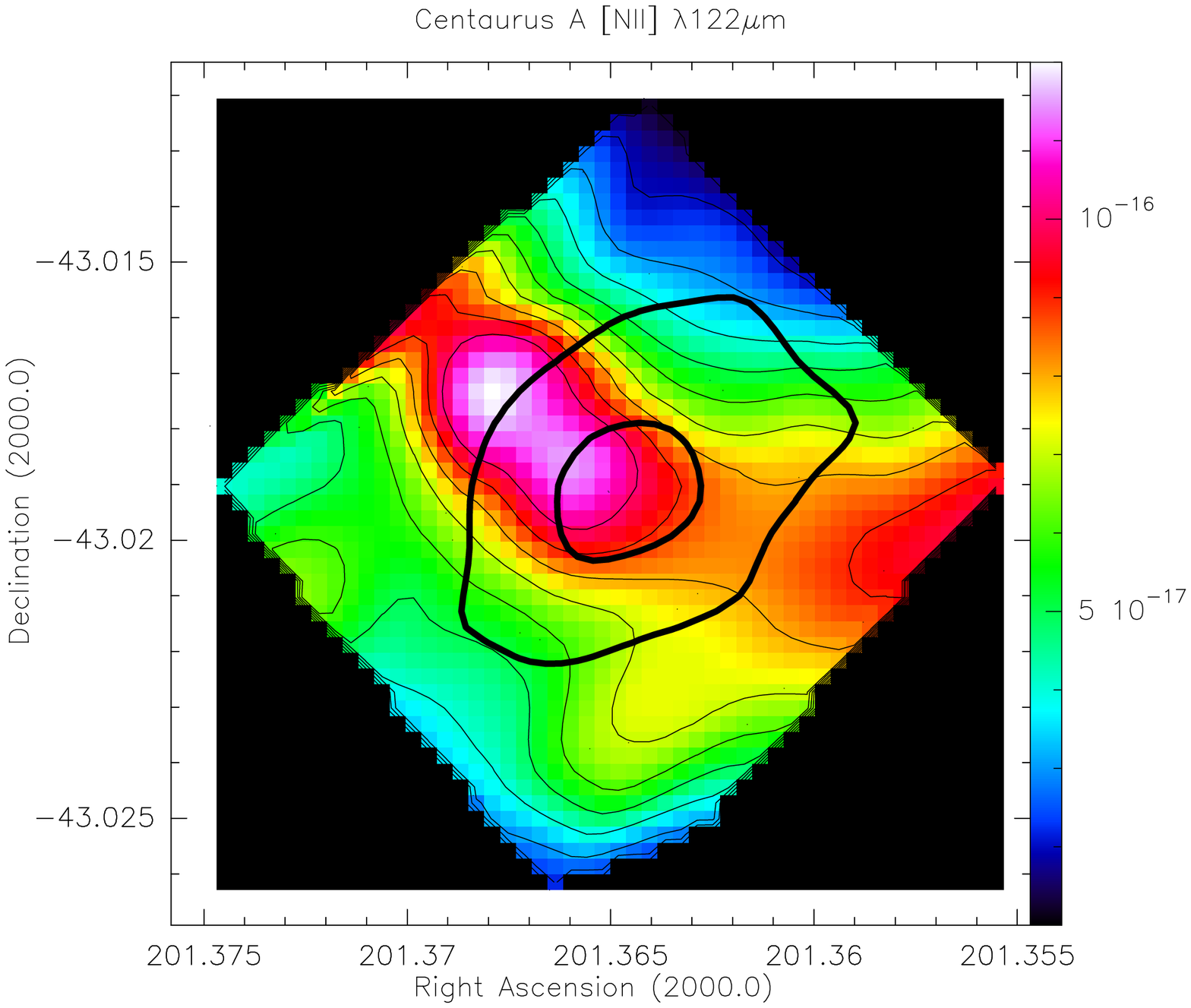}}}
\end{minipage}
\end{flushright}
\hspace{0.35cm}
\begin{minipage}[t]{18cm}
\begin{minipage}[t]{11.9cm}
\resizebox{6.8cm}{!}{\rotatebox{270}{\includegraphics*{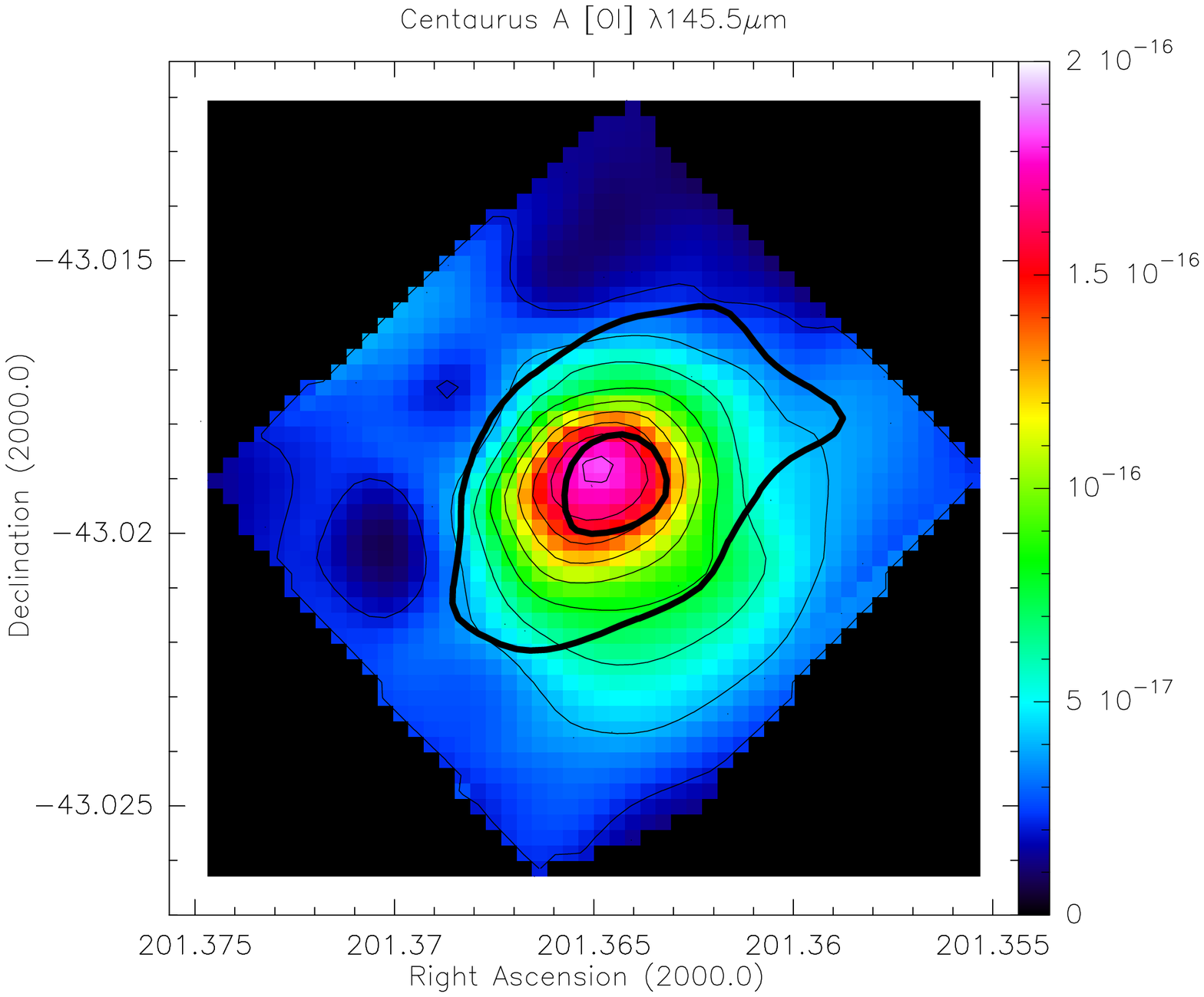}}}
\resizebox{5.55cm}{!}{\rotatebox{270}{\includegraphics*{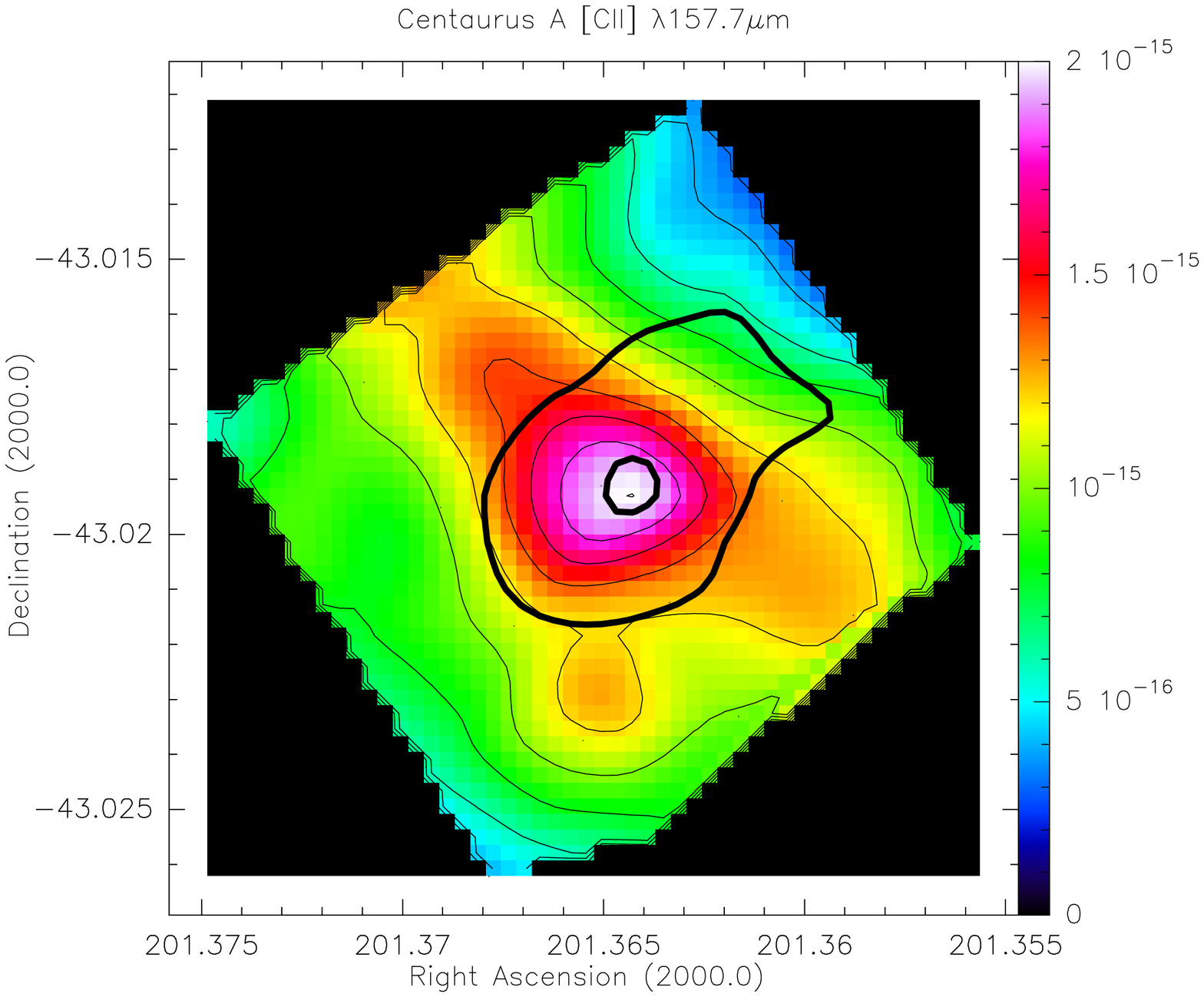}}}
\end{minipage}
\hfill
\hspace{0.4cm}
\begin{minipage}[t]{5.33cm}
\vspace{-0.01cm}
\resizebox{5.23cm}{!}{\rotatebox{0}{\includegraphics*{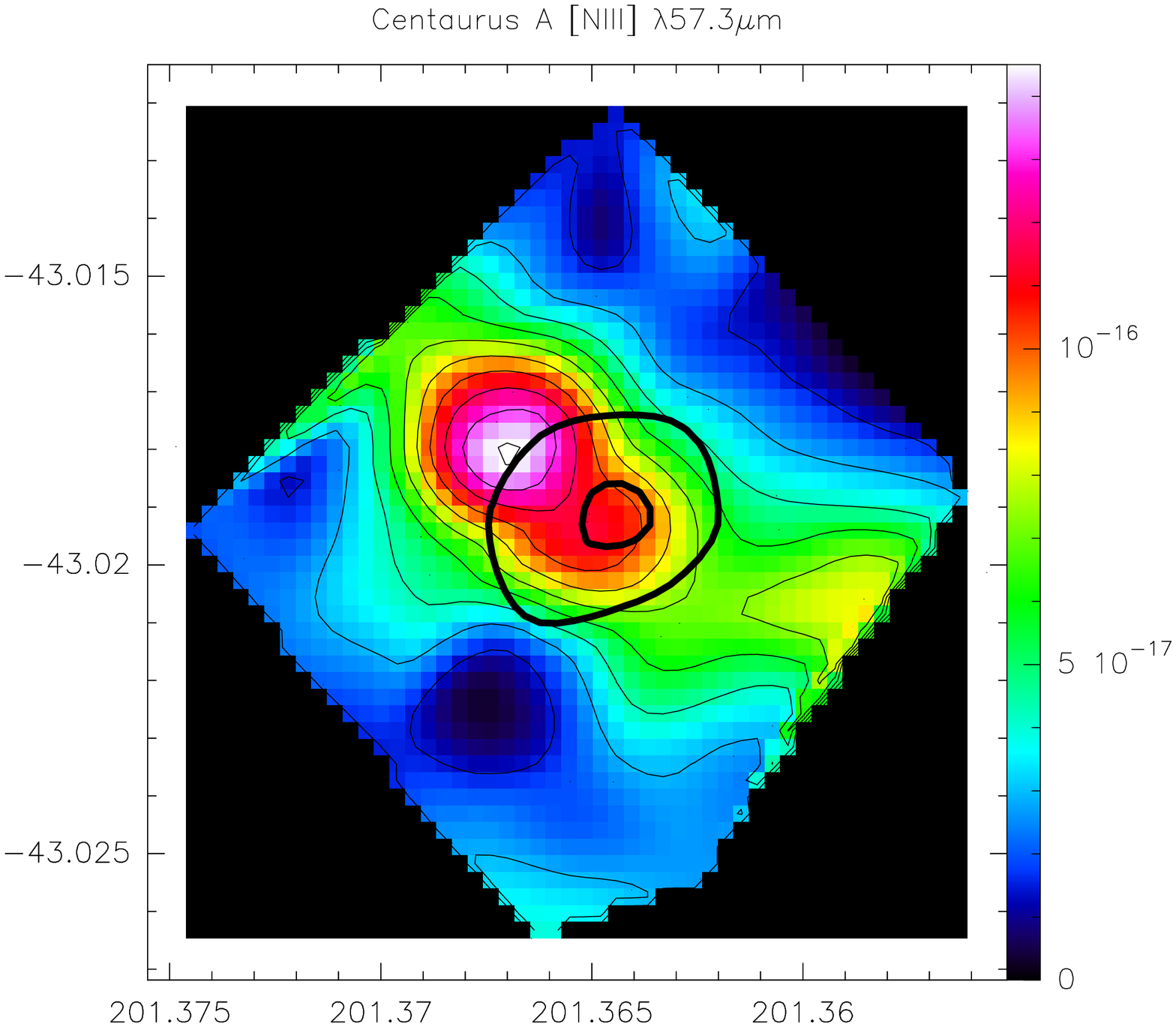}}}
\end{minipage}
\end{minipage}
\caption[]{Maps of the far-infrared fine structure line emission from
  the NGC~5128 central region. Left column: $\oi$63$\mu$m (top) and
  $\oi$145$\mu$m (bottom). Line emission contour levels are in steps
  of $4\times10^{-16}$ and $2\times10^{-17}$ W m$^{-2}$ respectively.
  Center column: $\cii$158$\mu$m (top) and $\oiii$88$\mu$m
  (bottom). Line emission contour levels are in steps of
  $3.5\times10^{-17}$ and $2\times10^{-16}$ W m$^{-2}$ respectively.
  Right column: $\nii$122$\mu$m (top) and $\niii$57$\mu$m
  (bottom). Line emission contour levels are in steps of
  $1\times10^{-17}$ and $1.5\times10^{-17}$ W m$^{-2}$ respectively.
  Thick solid lines mark the contours of the continuum emission at the
  line wavelength. Contours are at 8 and 16 Jy for the $\oiii$ and $\cii$ lines, and at 7 and 14 Jy for all other lines. 
  }
\label{linemaps}
\end{figure*}

\begin{table*}
\scriptsize
\caption{NGC~5128 spectral line map fluxes} 
\label{linedata}
\begin{center}                          
\begin{tabular}{lrrrrrrrr}        
\hline\hline                 
\noalign{\smallskip}
&\multicolumn{1}{c}{\niii}
&\multicolumn{1}{c}{\oi}
&\multicolumn{1}{c}{\oiii}
&\multicolumn{1}{c}{\nii}
&\multicolumn{1}{c}{\oi}
&\multicolumn{1}{c}{\cii} 
&\multicolumn{1}{c}{\nii$^a$}
&\multicolumn{1}{c}{\nii$^b$}\\
&\multicolumn{1}{c}{57$\mu$m} 
&\multicolumn{1}{c}{63$\mu$m} 
&\multicolumn{1}{c}{88$\mu$m} 
&\multicolumn{1}{c}{122$\mu$m}
&\multicolumn{1}{c}{145$\mu$m} 
&\multicolumn{1}{c}{157$\mu$m}
&\multicolumn{2}{c}{205$\mu$m}\\
\noalign{\smallskip}
\hline
\noalign{\smallskip}
spaxel & \multicolumn{6}{c}{Integrated line flux ($10^{-17}$ W m$^{-2}$]}\\
\noalign{\smallskip}
\hline
\noalign{\smallskip}
(0,0) &  1.88 (0.63) & 21.84 (4.09) &  8.97 (1.10) &  3.97 (0.41) &  2.96 (0.59) &  56.50 (0.55) &   0.33 (0.45) & 1.9 (2.6)\\
(0,1) &  2.52 (0.58) & 30.78 (2.52) & 11.70 (0.82) &  5.42 (0.44) &  2.60 (0.29) & 104.23 (0.37) &   1.00 (0.48) & 6.2 (2.8)\\
(0,2) &  6.62 (0.81) & 45.12 (2.17) & 24.12 (1.12) &  9.17 (0.50) &  3.99 (0.38) & 123.69 (0.43) &   0.86 (0.26) & 5.0 (1.5)\\
(0,3) &  1.84 (0.71) & 13.39 (2.05) &  6.55 (0.92) &  3.92 (0.28) &  2.58 (0.38) &  82.37 (0.37) &   0.93 (0.52) & 5.4 (3.0)\\
(0,4) &  1.37 (0.58) &  8.20 (3.19) &  2.39 (1.53) &  1.14 (0.27) &  0.55 (0.41) &  35.17 (0.44) &   0.40 (0.52) & 2.3 (3.0)\\
(1,0) &  2.33 (0.70) & 31.10 (2.65) & 12.81 (1.37) &  6.58 (0.31) &  3.44 (0.41) &  91.62 (0.37) &   0.53 (0.53) & 3.1 (3.1)\\
(1,1) &  4.82 (0.88) & 48.34 (3.97) & 12.84 (0.85) &  6.51 (0.44) &  3.39 (0.43) & 101.17 (0.64) &   0.75 (0.59) & 4.4 (3.4)\\
(1,2) & 13.55 (1.23) & 53.62 (2.64) & 26.98 (0.95) & 11.58 (0.39) &  5.12 (0.31) & 139.97 (0.96) &   1.94 (0.44) &11.3 (2.6)\\
(1,3) &  2.42 (0.44) & 21.98 (3.20) &  6.61 (1.38) &  3.83 (0.45) &  1.70 (0.35) &  65.41 (0.44) &   0.60 (0.35) & 3.5 (2.0)\\
(1,4) &  3.42 (0.60) &  7.89 (3.45) &  2.07 (1.09) &  1.63 (0.39) &  1.11 (0.41) &  27.60 (0.52) &   0.33 (0.45) & 1.9 (2.6)\\
(2,0) &  1.45 (0.58) & 25.94 (2.93) & 17.29 (1.32) &  3.76 (0.45) &  3.47 (0.56) &  73.93 (0.71) &   0.32 (0.32) & 1.9 (1.9)\\
(2,1) &  0.99 (0.49) & 64.18 (2.57) & 24.65 (0.89) &  5.75 (0.44) &  9.49 (0.66) & 110.25 (1.13) &   1.00 (0.37) & 5.8 (2.2)\\
(2,2) & 11.30 (0.99) &368.49 (2.75) & 35.44 (1.63) & 10.58 (0.49) & 23.49 (0.90) & 195.47 (1.30) &   1.74 (0.50) &10.1 (2.9)\\
(2,3) &  3.22 (0.64) & 54.38 (2.11) & 13.44 (1.10) &  5.46 (0.45) &  5.78 (0.36) & 102.19 (0.88) &   0.40 (0.60) & 2.3 (3.5)\\
(2,4) &  0.59 (0.58) & 12.84 (3.02) &  4.69 (0.91) &  2.07 (0.44) &  2.20 (0.34) &  36.24 (0.51) &   0.53 (0.59) & 3.1 (3.4)\\
(3,0) &  3.74 (2.05) & 23.97 (2.82) & 10.75 (1.08) &  3.33 (0.45) &  2.12 (0.40) &  60.11 (0.56) &   0.33 (0.61) & 1.9 (3.6)\\
(3,1) &  4.83 (2.19) & 41.87 (2.15) & 17.10 (0.75) &  6.98 (0.85) &  5.49 (0.89) & 114.82 (0.99) &   1.00 (0.52) & 5.8 (3.0)\\
(3,2) &  6.92 (0.95) & 76.77 (2.63) & 14.00 (1.17) &  7.84 (0.47) &  7.52 (0.45) & 124.04 (0.95) &   0.86 (0.30) & 5.0 (1.8)\\
(3,3) &  4.27 (0.94) & 47.89 (2.43) & 15.91 (1.01) &  7.96 (0.54) &  4.09 (0.53) & 126.06 (0.95) &   1.24 (0.46) & 7.2 (2.8)\\
(3,4) &  0.35 (0.61) & 27.52 (2.26) & 14.32 (0.85) &  5.65 (0.47) &  3.00 (0.42) &  86.33 (0.70) &   0.66 (0.45) & 3.8 (2.7)\\
(4,0) &  spike       &  7.05 (1.34) &  7.78 (1.58) &  2.00 (0.42) &  1.18 (0.27) &  38.65 (0.51) &   0.00 (0.00) & 0.0 (0.0)\\
(4,1) &  3.20 (1.81) & 16.20 (2.32) & 10.75 (1.08) &  2.85 (0.45) &  2.04 (0.33) &  70.85 (0.72) &   0.66 (0.45) & 3.8 (2.7)\\
(4,2) &  3.15 (0.78) & 34.26 (2.23) & 17.46 (0.67) &  7.11 (0.51) &  3.15 (0.34) &  97.67 (0.96) &   1.01 (0.41) & 5.9 (2.5)\\
(4,3) &  8.30 (1.42) & 36.38 (5.01) & 10.75 (1.08) &  9.26 (0.24) &  4.13 (0.29) & 123.57 (1.29) &   1.33 (0.34) & 7.7 (2.0)\\
(4,4) &  2.66 (0.66) & 43.25 (2.69) & 24.40 (1.24) &  9.39 (0.29) &  2.25 (0.29) &  73.55 (0.86) &   0.56 (0.36) & 3.3 (2.2)\\
\noalign{\smallskip}
\hline
\end{tabular}
\end{center}
Notes: a: Flux from nominal PACS calibration. b: Flux scaled to SPIRE calibration (see text).
\end{table*}

\begin{table*}
\scriptsize
\caption[]{Overview of atomic line fluxes}
\begin{center}
\begin{tabular}{lccccccccc}
  \noalign{\smallskip}     
  \hline\hline
  \noalign{\smallskip}
Line & $ISO^{a}$&$PACS5^{b}$&$PACS3^{c}$&$PACS1^{e}$& Nucleus$^{f}$&Northern  & $PACS$ & $ISO-PACS3$\\
& LWS     &  Sum      &    Sum   &          &              & Jet$^{f}$ &mean$^{g}$ & mean$^{h}$\\
& $70"$   &$47"x47"$  &$28"x28"$ &$9.4"x9.4"$&  P          &  P       &           &  \\
& \multicolumn{6}{c}{---------------------------------------- ($10^{-16}$ W m$^{-2}$) ----------------------------------------}&\multicolumn{2}{c}{--------- (10$^{-7}$ W m$^{-2}$ sr$^{-1}$) ---------}\\
\noalign{\smallskip}     
\hline
\noalign{\smallskip} 
$\oiii$(52$\mu$m) &  72  &   --   &  --  &  --   &  --  & --:  &  --  & --  \\
$\niii$(57$\mu$m) &  24  &   9.7  &  5.2 &  1.1  &  1.4 & 1.8  & 0.2  & 0.3 \\
$\oi$ (63$\mu$m)  & 196  &  116   &  95  & 37    &  52  &  3.3 & 1.1  & 1.4 \\
$\oiii$ (88$\mu$m)&  70  &   35   &  25  &  3.5  &  4.8 &  3.5 & 0.5  & 0.6 \\
$\nii$ (122$\mu$m)&  15  &   15   &  10  &  1.1  &  1.5 &  1.7 & 0.2  & 0.07\\ 
$\oi$ (145$\mu$m) &  11  &   10   &  8.5 &  2.3  &  4.3 &$<$0.5& 0.1  & 0.04\\
$\cii$ (158$\mu$m)& 291  &  226   & 108  & 19.5  &  34  & 20   & 3.3  &1.7 \\
$\nii$(205$\mu$m)$^{i}$ &--& 1.9  &  1.0 &  0.17 & 0.25  &  0.3 & 0.03 & -- \\
                  &  --  &  11    &  5.6 &  1.0  &  1.5 &  1.9 & 0.18 & -- \\
FIR (43-197$\mu$m)&80000 & 41500  & 24150& 7195  & 8000 & 2500 & 540  & 780 \\
$\ci$ (370$\mu$m)$^{k,d}$&--& 5.72 & 2.68 & 1.27  &  --  &  --  & 0.14 & -- \\
$\ci$ (610$\mu$m)$^{k,d}$&--& 1.70 & 0.90 &  0.43 &  --  & 0.21 & 0.04 & -- \\
CO(4-3) (651$\mu$m)$^{k,d}$&--&1.72&0.69& 0.31 &  --  & 0.13 & 0.04 & -- \\
\noalign{\smallskip}     
\hline
\noalign{\smallskip}
\end{tabular}
\end{center}
Notes: 
(a) Unger \etal (2000); area $9.04\times10^{-8}$ sr;
(b) Sum of all fluxes in the $5\times5$ spaxels of the PACS array; 
area $5.64\times10^{-8}$ sr;
(c) Sum all fluxes in the $3\times3$ inner spaxels of the PACS array; 
area $1.87\times10^{-8}$ sr;
(d) taken or extrapolated from data in Israel \etal (2014); 
(e) Flux in central PACS spaxel, assuming isolated point source;
(f) Assumed to be a point source, corrected for emission from adjacent
spaxels;
(g) mean surface brightness in PACS map outside CND and jet regions;
area $4.36\times10^{-8}$ sr;
(h) mean surface brightness over {\it ISO} LWS beam area excluding the 
bright inner PACS $3\times3$ spaxel region; area $7.18\times10^{-8}$ sr;
(i) first line: nominal PACS calibration; second line: scaled to SPIRE calibration) 
(k) APEX data
\label{alllinedata}
\end{table*}

The nuclear point source outshines the entire CND at all wavelengths
shorter than 25$\mu$m and longer than $350\mu$m.  The
60$\mu$m/160$\mu$m flux density ratios of the CND and ETD emission are
significantly different at values of 0.95 and 0.55 respectively. This
implies that the CND is warmer than the ETD, with $T_{d}$=31 K versus
$T_{d}$=28 K for a Rayleigh-Jeans dust spectral index $\alpha$ = -4,
close to the values found by Parkin \etal (2012) for the ETD emission
from Cen~A.  The CND has a far-infrared luminosity ({\it IRAS}
definition) $FIR\,=\,1.59\times10^{-12}$ W m$^{-2}$. If the dust
composition and size distribution in the CND are similar to those of
the Milky Way - an uncertain assumption, see Galliano \etal 2011 - the
CND dust mass would be $M_{dust}\,\approx\,3.5\times10^{5}$
M$_{\odot}$ with a formal uncertainty less than $10\%$. In Paper I, we
found a CND gas mass $M_{gas}\,\approx\,8.4\times10^{7}$ M$_{\odot}$
within a factor of two (Israel \etal 2014) implying a CND gas-to-dust
ratio of 240 with a similar uncertainty.  In section 5.4 of this
paper, we will refine the mass determination to
$M_{gas}\,=\,9.1\pm0.9$ M$_{\odot}$, changing the gas-to-dust ratio to
$260\pm40$.  This value falls within the range of gas-to-dust ratios
100-300 found by Parkin \etal (2012) for the ETD, but neither takes
into account the systematic uncertainty caused by the assumed dust
properties.

Finally, we note that in AGNs, the presence of a central compact
far-infrared source is often interpreted as a high-star-formation
nuclear cusp (see \eg Mushotzky $\etal$ 2014). The example of Cen~A
shows that this is not necessarily the case. The CND is compact and
bright in the far-infrared, but its molecular gas is cold and not
associated with star formation (Paper I). We could establish its
true nature only because it is so near.

\subsection{The submillimeter emission lines}

The unusually high brightness of the \ci\ lines from the Cen~A center
with respect to the CO ladder is further illustrated by our new APEX
$\ci$ 492 GHz and $\co$ $J$=4-3 maps. As discussed in Paper I,
emission profiles from the central position are affected by line
absorption against the nuclear continuum point source.  To correct for
this, we have determined velocity-integrated temperatures
$\int T_{mb}$d$v$ from gaussians fitted to the observed profiles,
excluding the velocity range $V_{LSR}$ = 520-620 $\kms$ affected by
absorption. The absorption-corrected CO(4-3) and $\ci$ maps are shown
in Fig.\,\ref{carbonmapfig}.  In both maps, the corrected emission
peaks at the nucleus within the pointing error.

The $\ci$ emission is more compact (deconvolved FWHM $11"$) than the
CO (4-3) emission (deconvolved FWHM $16"$).  The bright $\ci$ is
asymmetrical and fans out to the northeast.  There is relatively
little $\ci$ emission from the ETD.  In contrast to the $\ci$, the
bright CO is remarkable for its symmetrical extension along the
northeast-southwest axis, in the same position angle as the radio
jet. Its extension along the CND major axis is less conspicuous. Low
surface brightness CO emission fills almost the entire map area
implying that the ETD contributes significantly to the total CO
emission.

In the rightmost panel of Fig.\,\ref{carbonmapfig}, the map of the
$\ci$ to CO (4-3) integrated brightness temperature ratio
reveals the relative strength of $\ci$ in both the CND and in the ISM
extending to its northeast.  Towards the CND, the $\ci$ line is $40\%$
stronger than the CO (4-3) line. Northeast of it the ratio is 1.2.
Thus, the unusually high $\ci$/CO (4-3) ratio noted in Paper I can be
directly related to the inner CND.  Away from the CND, the ratio drops
to the value of 0.5 that characterizes the ETD ISM at greater
distances from the nucleus.  This ratio is practically identical to
the mean value of $0.46\pm0.03$ in a sample of actively star-forming
galaxy centers (Israel \etal 2015) and it is thus consistent with the
vigorous star-formation seen to take place in the ETD.

The differences between the $\ci$ and CO (4-3) distributions are
underlined by the position-velocity ($pV$) diagrams in
Fig.\,\ref{pvplots} which are constructed from the observed
profiles, not corrected for absorption against the nucleus.
The $pV$ diagrams along the CND major axis (lefthand panels) are
almost complementary. In the $\ci$ diagram (top left), the bright and
well-defined CND emission contrasts strongly with much fainter ETD
emission, once again underlining the dominating nature of $\ci$ in the
CND. The CO (4-3) $pV$ diagram (bottom left) shows the opposite: the
bright CO emission from the ETD outshines the fainter CO in the CND.

Especially intriguing are the $pV$ diagrams along the
northeast-southwest axis perpendicular to the CND.  The $\ci$ $pV$
diagram (top right) reveals bright CND emission over the full velocity
width of $\Delta V$ = 200 $\kms$, the CO $pV$ diagram (bottom right)
shows the bright emission to be more concentrated in velocity.  We
draw attention to the extensions perpendicular to the CND, at $Y$ =
-15$"$, $V_{\rm LSR}\,=\,580,\kms$, and at $Y$ = +20$"$,
$V_{\rm LSR}\,=\,480\,\kms$ which are relatively faint in the $\ci$
diagram, but more prominent in the CO (4-3) diagram. These features
cover more than twice the CND extent in $Y$; they are not part of it.
The observed CO $pV$ distribution has the appearance of a blend of two
components overlapping in the center, each at its own discrete
velocity, most likely the signature of a bipolar outflow in both $\ci$
and CO. Unfortunately, the APEX resolution is insufficient to
determine more detail that would provide further clues to the nature
of the suspected outflow, and the relative roles of $\ci$ and CO.
 
Table\,4 in Paper I provided the central $\ci$ fluxes in both the
$J$=2-1 and the $J$=1-0 line in different apertures. With the aid of
the $\ci$ (1-0) map in Fig.\,\ref{carbonmapfig} we have reduced these
measurements to common apertures for both lines. We found a constant
$I_{\ci(2-1)}/I_{\ci(1-0)}$ ratio of $0.67\pm0.05$ (corresponding to a
flux ratio of $3.0\pm0.3$) for apertures ranging from $10"$ to $28"$.

\subsection{The far-infrared fine-structure lines}

The PACS spectral line maps (Fig.\,\ref{linemaps}) show emission
throughout the $47"\times47"$ ($0.9\times0.9$ kpc) region mapped,
although intensities approach zero at the northern boundary.  Much of
this emission must be due to the ETD, but at the PACS resolution,
details are washed out and there is no clear counterpart to the
`parallelogram' structure seen in other maps (cf. Espada \etal
2009). In addition to the extended diffuse emission, all line maps
show compact, bright emission coincident with the center of the Cen~A
CND, as well as the northern X-ray/radio jet (cf. the $\cii$, $\nii$,
$\niii$, and $\oiii$ line maps). Although the CND is very clearly
outlined in the continuum maps, the line maps do not unambiguously
reveal CND emission outside the center. 

The resolution of the images in Fig.\,\ref{linemaps} is insufficient
to determine whether the strong central emission seen in the $\oi$ and
$\cii$ lines represents a true point source exclusively associated
with the nuclear region, or a slightly extended source incorporating
the inner CND.  We will address this question in Section 5.2. Nitrogen
line emission is seen primarily towards the northern jet, peaking at a
projected distance of $9"$ (165 pc) from the nucleus. This is unlike
the emission in the $\cii$ and $\oiii$ lines that also extends towards
the northern jet but peaks at the center.  The ionized nitrogen and
carbon maps also show a weak extension to the southwest, away from the
jet direction.

In Table\,\ref{alllinedata}, we have listed the line fluxes from the
various parts of the PACS maps together with the {\it ISO} LWS fluxes
from Unger \etal (2000).  We have included the $\ci$ and CO(4-3)
fluxes from the APEX maps, where necessary integrated over an area
corresponding to the PACS flux sums. Comparison of the line
intensities in Table\,\ref{alllinedata} shows that, notwithstanding
the bright emission from the Cen~A nucleus and jet positions, $\cii$,
$\oi$ and $\oiii$ emission is also widespread in the ETD.  The weak
$\oi$145$\mu$m line contributes proportionally less to the ETD than
the much stronger $\oi$63$\mu$m line.  The $\nii$ line emission is
seen predominantly towards the nucleus and the jet.

\section{Analysis}

\subsection{LVG modeling of the neutral gas from the
  sub-millimeter lines}

\begin{figure}[]
\unitlength1cm
\begin{center}
\resizebox{6.6cm}{!}{\rotatebox{270}{\includegraphics*{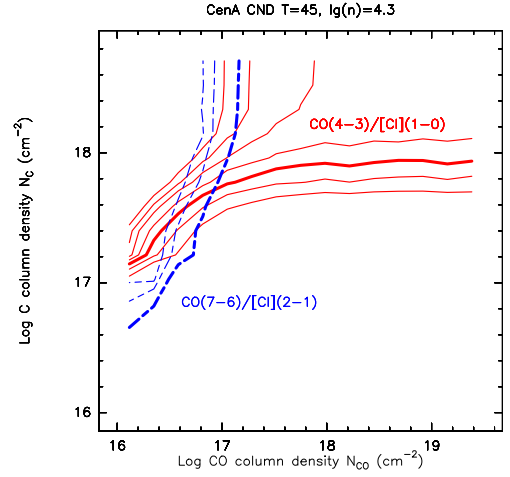}}}
\end{center}
\caption[]{
  With the $\h2$ temperature and density defined by the intersection
  of the $\ci$(2-1)/(1-0) and $\co$(7-6)/(4-3) curves, the flux ratios
  $\ci$(1-0)/$\co$(4-3) (red) and $\ci$(2-1)/$\co$(7-6) (blue) can be
  used to find neutral carbon and carbon monoxide column
  densities. Curves are shown for $\co$(4-3)/$\ci$(1-0) ratios of 1.2
  to 0.6 (left to right; the flux ratio of 0.8 is marked by a thicker
  line) and for $\co$(7-6)/$\ci$(2-1) ratios of 8, 6, and 4 (thick
  line). }
\label{cencico}
\end{figure}

In a previous paper (Israel $\etal$ 2015), we presented an LVG
analysis using the Leiden RADEX code of the average properties of the
molecular ISM in a sample of 76 luminous galaxies based on the $\ci$ and
selected CO line fluxes.  Cen~A is part of the sample but its line
ratios differ significantly from those of the other galaxies at
similar distances.  Here we have again applied the method described in
that paper to determine the (molecular) ISM properties of the Cen-A
CND.  We have reduced the fluxes from Paper I and this paper to an
aperture of $19"$, essentially covering the CND.  The combined CO and 
\ci\ line ratios 
define a mean CND temperature
$T_{kin}$ = 45 K and a density log $n(\h2)$ = 4.2 $\cm2$.  The
CO/$\ci$ ratios Fig\,\ref{cencico} (bottom) was then used to
determine the CND mean CO and $\ci$ column densities at this
temperature and density.  The various $\co$(4-3)/$\ci$(1-0) curves
correspond to the range of ratios evident in the map in
Fig.\,\ref{carbonmapfig}; the three curves shown for
$\co$(7-6)/$\ci$(2-1) illustrate that the results are relatively
insensitive to the actual value of that ratio.

The interpretation of the observed fluxes in model terms is
complicated by the possibility that part of the $\ci$ emission from
the center is not due to the CND, but comes from the outflow, and that
the actual CND temperature is lower than 45 K.  We have therefore
derived three different solutions exploring the possible range of CND
physical parameters. The three solutions are summarized in
Table\,\ref{cicotab}.  The first assumes that the nominal line ratios
at the center are representative for the CND. The second assumes that
only two thirds of the central $\ci$ flux comes from the CND, but that
the CND temperature is unchanged at 45 K. The third solution assumes
that again only two thirds of the flux comes from the CND, and that
the temperature is only 25 K.

\begin{table}
\scriptsize
\caption[]{Cen~A center dense$^{a}$ neutral gas amounts$^{b}$}
\begin{center}
\begin{tabular}{lccc}
\hline
\noalign{\smallskip}
Case$^{c}$                           &Nominal & Reduced &Reduced\\
Temperature                         & 45 K   &   45 K  & 25 K  \\
\noalign{\smallskip}     
\hline
\noalign{\smallskip}
Model $N$(CO)/d$V$ (10$^{17}\kkms\cm2$)& 0.85   & 1.0   &  3.0    \\  
Model $N$(CI)/d$V$ (10$^{17}\kkms\cm2$)& 10     & 5.4   &  4.0    \\
Mean $N$(CO) (10$^{18}\cm2$)         & 0.19   & 0.19  &  1.1    \\
Mean $N$(CI) (10$^{18}\cm2$)         & 1.7    & 0.83  &  1.3    \\
$M_{densegas}$(CO)$^{d}$ (10$^{7}$ M$_{\odot}$) & 0.3    & 0.3   & 1.85    \\
$M_{densegas}$(CI)$^{d}$ (10$^{7}$ M$_{\odot}$) & 2.8    & 1.35  & 2.15    \\
$M_{tenuousgas}$(CO)$^{d}$ (10$^{7}$ M$_{\odot}$) & 1.0    & 1.0   & 1.0    \\
\noalign{\smallskip}     
\hline
\end{tabular}
\end{center}
Note: a. All values for log n($\h2$)=4.2; 
b. Uncertainties in the derived column densities and masses are 
   typically $50\%$
c. See text, section 4.1
d. Includes a contribution by helium of $35\%$ by mass
\label{cicotab}
\end{table}

From the model column densities, the corresponding gas masses can be
calculated. We assumed a CND surface area of $6.8\times10^{-9}$ sr
($1\times10^{5}$ pc$^{2}$), and applied filling factors defined as the
ratio of the observed flux to the model flux.  In order to transform
carbon column densities into those of total hydrogen, we assumed a
metallicity of 0.75 solar (Table\,\ref{lineratios}), i.e 12+log[O]/[H]
= 8.55, from which we deduce log [C]/[O] = $-$0.35 (cf. Garnett \etal
2004). This defined a carbon elemental abundance
$x_{C}\,=\,1.6\,\times\,10^{-4}$. In the neutral gas, a significant
fraction of all carbon may be locked up in dust grains. Reliable
estimates of the resulting carbon elemental depletion exist only for
the Milky Way (Jenkins, 2009). Following this work, we assumed that
$60(\pm20)\%$ of all carbon is locked up in dust.  Hence, we found a
carbon neutral gas-phase abundance
$x'_{C}\,=\,0.65\,\times\,10^{-4}$. Finally, all masses were multiplied
by a factor of 1.35 to account for the helium component of the ISM.
The resulting masses are given in Table\,\ref{cicotab}.  In all cases,
there is more neutral carbon than carbon monoxide.  Only if we lowered
the temperature to 25 K, does the inferred amount of carbon in neutral
atomic form approach that in molecular form.

However, this is not all. The line ratios we have used so far are
relatively insensitive to molecular gas at low temperatures and
densities, that is mostly sampled by $\co$ emission in lower $J$
transitions.  With RADEX, we have modeled these lower $J$ CO
lines in two extreme cases. In the first, we assumed that this cool,
tenuous gas does not contribute to the $J$=4-3 transition. In this
case, we found a mass $M_{gas}=(1.5\pm0.7)\times10^{7}$ M$_{\odot}$
with kinetic temperature $T_{kin}=20$ K and space density
$n_{\h2}=100\,\cc$.  This would significantly increase the CO masses
in Table\,\ref{cicotab}, making CO a much more important component. In
the second case, we allowed as much as half of the CO(4-3) flux to come
from the tenuous gas, which then required a lower additional mass
$M_{gas}=(0.5\pm0.3)\times10^{7}$ M$_{\odot}$ at a substantially
higher kinetic temperature $T_{kin}=125\pm25$ K and higher space
density $n_{\h2}=500\,\cc$. The previously derived dense gas
space density is increased by a factor of a few but its mass is
virtually unchanged, and the impact of the additional mass to the the
total is modest.  In either case, the amount of gas associated with
neutral carbon is essentially unchanged.  We thus conclude that there is
an additional mass $M_{gas}=(1.0\pm0.5)\times10^{7}$ M$_{\odot}$
associated with tenuous molecular gas over and above the mass of dense
gas deduced from Fig.\,\ref{cencico}, also listed in
Table\,\ref{cicotab}. 

In Paper I, we presented three possible solutions to the LVG analysis
of the CND $\co$ and $\thirco$ fluxes but were unable to decide which
of these was most representative of actual conditions. The analysis
presented here suggests the final of these three solutions best
describes the CND molecular gas content.

\subsection{PDR modeling of the neutral and ionized gas from the
  far-infrared lines}

\begin{figure*}[]
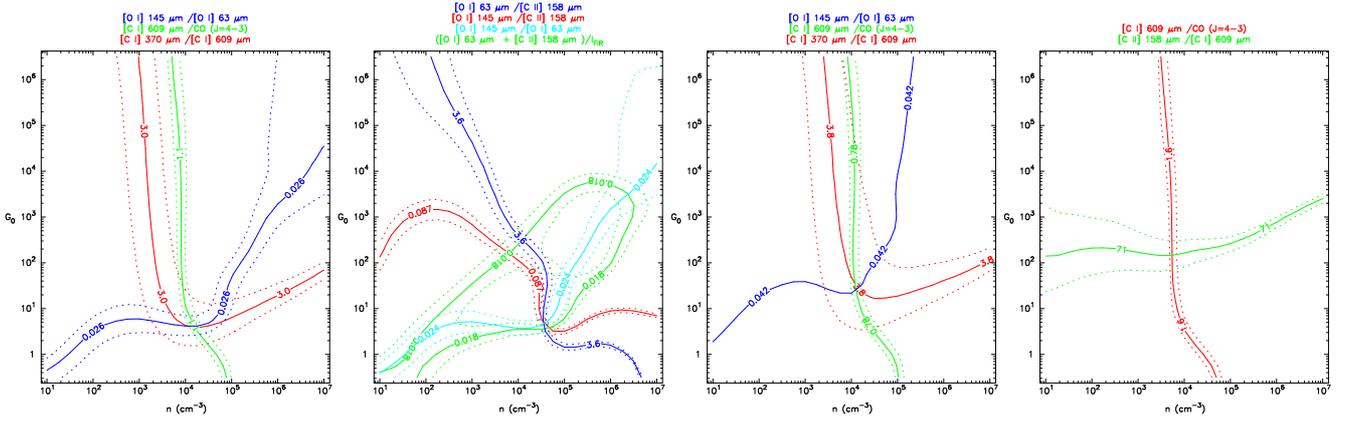

\unitlength1cm
\begin{minipage}[t]{18cm}
\resizebox{4.3cm}{!}{\rotatebox{0}{\includegraphics*{CENA_CND1.eps}}}
\resizebox{4.3cm}{!}{\rotatebox{0}{\includegraphics*{CENA_CND2.eps}}}
\resizebox{4.3cm}{!}{\rotatebox{0}{\includegraphics*{CENA_ETD1.eps}}}
\resizebox{4.3cm}{!}{\rotatebox{0}{\includegraphics*{CENA_NJET1.eps}}}
\end{minipage}
\caption[]{Determination of radiation field $G$ and density $n_{\rm o}$ 
from PDR Tool Box line ratio fits for the CND (left two panels), ETD 
(second from right), and outflow (rightmost) components using the data 
from Table\,\ref{lineratios}. The two CND panels illustrate the sequential 
procedure: in the lefthand CND diagram we fit the $\oi$ line ratio to the 
intersection of the $\ci$ and the $\ci$/CO line ratios in order to find 
the $\oi$ absorption multiplication factor and then plot the corrected 
$\oi$ and $\cii$ line ratios in the diagram next to it. }
\label{toolboxfit}
\end{figure*}

From the fluxes in Table\,\ref{alllinedata}, we have derived the
representative line ratios in Table\,\ref{lineratios}.  These line
ratios were then used to diagnose the physical parameters also listed
in that table.

\begin{table}
\scriptsize
\caption[]{Diagnostic line ratios and physical parameters}
\begin{center}
\begin{tabular}{lcccc}
\noalign{\smallskip}     
\hline\hline
\noalign{\smallskip}
Ratio                           &ETD$^{a}$&CND$^{b}$&Nucleus$^{c}$&Outflow$^{d}$ \\
\noalign{\smallskip}     
\hline
\noalign{\smallskip} 
\multicolumn{5}{c}{\bf Ionised gas (HII regions)}\\
\niii57$\mu$m/\nii122$\mu$m   & 0.9    & 0.3    & 1.0   & 1.1  \\ 
\niii57$\mu$m/\oiii88$\mu$m   & 0.45   & 0.1    & 0.3   & 0.5  \\ 
\nii122$\mu$m/\nii205$\mu$m$^{e}$& 0.9  & 2.9    & 1.0   & 0.9  \\
\oiii88$\mu$m/\nii122$\mu$m   & 2      & 2.5    & 3.2   & 2.1  \\
\oiii88$\mu$m/\oi63$\mu$m     & 0.5    & 0.4    & 0.09  & 1.1  \\
\cii158$\mu$m/\nii122$\mu$m   & 24     & 8.8    & 21    & 12   \\
\cii158$\mu$m/\nii205$\mu$m$^{e}$&22   & 25      & 22   & 11   \\
\noalign{\smallskip}     
\hline
\noalign{\smallskip} 
Electron density $n_{e}\, (\cc)$ & 10    & 100   & 15     & 10 \\
Stellar temperature $T_{eff}$ (K)& 34300 & 32300 & 34500  & 35000\\
Per centage \cii\ from ionized zone& 7.5 &  10   &  12   & 25  \\
Metallicity $Z$ (Z$_{\odot})$     & 0.75   & 0.7   &  0.8  & 1 \\
\noalign{\smallskip}     
\hline
\noalign{\smallskip} 
\multicolumn{5}{c}{\bf Neutral gas (PDRs)}\\
\ci370$\mu$m/\ci609$\mu$m     & 3.8    & 3.0        &  3.0        & ---    \\
\ci609$\mu$m/CO(4-3)          & 0.8    & 1.2        &  1.4        & 1.6    \\
\oi63$\mu$m/\oi145$\mu$m      & 16     & 10         & 13          & $>6$   \\
\oi63$\mu$m/\cii158$\mu$m     & 0.36   & 0.62       & 1.65        & 0.17   \\
\oi63$\mu$m+\cii158$\mu$m/FIR & 0.0040 & 0.0040     & 0.0095      & 0.0090 \\
\cii158$\mu$m/FIR             & 0.0025 & 0.0030     & 0.0035      & 0.0070 \\
\noalign{\smallskip}     
\hline
\noalign{\smallskip}
PDR gas density log $n_{\rm o}\,(\cc)$&3.5$\pm$0.7& 4.3$\pm$0.2& 4.4$\pm$0.3& 3.6$\pm0.21$\\
PDR radiation field $G\,(\go)$      & 40$\pm$10  & 4.0$\pm$0.5& 4.5$\pm$0.5& 130$\pm$60\\
PDR surface temperature $T_{s}$ (K)  & 70$\pm$10 & 50$\pm$5   & 40$\pm$5   & 160$\pm$20 \\
$\tau_{OI}$                          &0.5$\pm$0.1 & 1.5$\pm$0.3& 1.0$\pm$0.2& --\\
\noalign{\smallskip}     
\hline
\noalign{\smallskip} 
\end{tabular}
\end{center}
Notes: 
(a) PACS5-PACS3 from Table\,\ref{alllinedata}); 
(b) PACS3-0.5(PACS1/Nucleus) from Table\,\ref{alllinedata}; 
(c) 0.5(PACS1/Nucleus) from Table\,\ref{alllinedata}; 
(d) Jet entries from Table\,\ref{alllinedata}; 
(e) Fluxes scaled to SPIRE calibration;
\label{lineratios}
\end{table}

For the ionized gas, we have used the $\oiii$, $\nii$, and $\niii$
lines that all have ionization potentials above 13.6 eV (see
Table\,\ref{targets}) guaranteeing that they originate in fully
ionized ($\hii$) regions. Electron densities $n_{e}$ follow from the
$\nii$122$\mu$m/$\nii$205$\mu$m line ratios, after which the
$\cii$158$\mu$m/$\nii$205$\mu$m is used to find the fractions of the
$\cii$ emission from neutral and fully ionized gas (cf. Oberst \etal
2006). The electron densities so deduced are low throughout
($n_{e}\,\approx\,10$ cm$^{-3}$) except in the CND where they are ten
times higher. Fully ionized gas is a modest contributor to the
observed $\cii$ emission, except in the outflow region where it
provides a quarter of the flux. The $\niii$57$\mu$m/$\nii$122$\mu$m
ratio gauges the hardness of the UV radiation field expressed as the
mean effective temperature T$_{eff}$ of the exciting stars (see Rubin
\etal 1994, Abel \etal 2005). It is lowest in the CND and highest in
the outflow, but the differences are small. The metallicity is derived
from the ratio of the $\oiii$ and $\nii$ lines to the $\niii$57$\mu$m
line (Nagao \etal 2012). The modestly sub-solar metallicity in
Table\,\ref{lineratios} is consistent with the value of $0.95\pm0.35$
derived from X-ray observations by Markowitz $\etal$ 2007, and the
finding by M\"ollenhof (1981) that the $\hii$ regions in the ETD have
roughly solar abundances.

We have used the on-line calculator PDR Tool Box described by Pound
$\&$ Wolfire (2008) to derive the conditions in the neutral gas. This
exploits the distinctly different dependencies on neutral gas density
$n_{\rm o}$ and irradiating UV field $G$ exhibited by the various line
ratios of species such as $\ci$, $\oi$, $\cii$ and CO combined with
the FIR continuum emission (Kaufman $\etal$ 1999; Kaufman, Wolfire
$\&$ Hollenbach 2006).

Although the bright $\oi$63$\mu$m line is commonly used in combination
with the $\cii$ line and the FIR continuum as the primary diagnostic
for PDR density and irradiation, its actual usefulness is severely
limited because it can be strongly absorbed by its foreground (see for
instance Vasta \etal 2010).  For this reason, we have first determined
the parameters $n_{\rm o}$ and $G$ from the CO (4-3) and the two $\ci$
lines only.  This procedure was feasible owing to the fact that both
$\ci$ lines have been observed, which is often not the case.  The
results were in excellent agreement with those obtained from the LVG
analysis in the previous section.  In the next step, we determined the
$\oi$63$\mu$m line intensity multiplication factor that makes the
$\oi$145$\mu$m/$\oi$63$\mu$m ratio fit those values.  We found optical
depths of $1.5\pm0.3$ for the CND emission and $0.54\pm0.10$ for the
ETD emission from the multiplication factors of 4.5 and 1.7 thus
obtained (Fig.\,\ref{toolboxfit} and Table\,\ref{lineratios}). An
independent determination of $n_{\rm o}$ and $G$ from the corrected
$\oi$ and $\cii$ line fluxes combined with the FIR continuum flux
density yielded consistent results. We emphasize that the use of $\oi$
line fluxes in the diagnostic diagrams would have led us to both
underestimate the density and overestimate the radiation field by more
than an order of magnitude if we had trusted the observed $\oi63\mu$
line flux without correction.

In the very center of the Cen~A CND, the atomic gas cooling is
dominated by the $\oi$63$\mu$m line even though the line is strongly
absorbed.  This is the only part of Cen~A where the cooling is not
dominated by the $\cii$158$\mu$m line.  The neutral gas densities are
much higher than those of the ionized gas, which suggests that the
latter are low-density outer layers of dense neutral clouds, ionized
by the UV photons from the luminous stars in the ETD. The radiation
field of the starburst-hosting ETD, $G\,\approx\,40$ G$_{\circ}$,
exceeds that of the CND by an order of magnitude, consistent with the
absence of star formation in the CND.  The very low
$\oiii$88$\mu$m/$\oi$63$\mu$m ratio further illustrates the dominance
of the $\oi$ emission from very dense neutral gas in the CND, in stark
contrast to the ten times higher ratio in the more ionized outflow.

Finally, the column density of ionized carbon can be calculated if
both gas density and kinetic temperature are known (Pineda \etal 2013,
Eqn 1). We used the data from Table\,\ref{alllinedata}, with a
spontaneous decay rate $A_{ul} = 2.3\times10^{-6}$, and collisional
de-excitation rate coefficients $R_{ul}$ (electrons) =
$1.4\times10^{-6}\,T^{-0.37}$, and $R_{ul}$ ($\h2$ molecules) =
$3.4\times10^{-10}\,T^{-0.02}$. Carbon column densities thus
obtained are transformed into the column densities of the parent
hydrogen gas by assuming the abundances discussed in Section 4.1. for
a metallicity of 0.75 solar.

\begin{table}
\scriptsize
\caption[]{Amount of $\cii$-related gas in the Cen~A center}
\begin{center}
\begin{tabular}{lccc}
\noalign{\smallskip}
\hline\hline
\noalign{\smallskip}
Case                      &\multicolumn{3}{c}{H mass traced by [CII] emission}\\
                          &outer CND  & central CND & Outflow  \\
                          &\multicolumn{3}{c}{$10^{7}$ M$_{\odot}$} \\
\noalign{\smallskip}     
\hline
\noalign{\smallskip} 
Surface area ($10^{-9}$ sr)&  4.7      & 2.1      & 2.1  \\ 
Limiting value$^{a}$       & 0.72      & 0.33     & 0.25 \\
Ionized hydrogen$^{b}$     & 0.08      & 0.08     & 0.17 \\
Neutral hydrogen$^{c}$     & 2.1       & 1.5      & 0.43 \\ 
Total gas $^{d}$           & 2.9       & 2.1      & 0.8 \\
\noalign{\smallskip}     
\hline
\end{tabular}
\end{center}
Notes: (a) High-density ($n>>3000\cc$), high-temperature ($T>>91$ K)
limit yielding minimum total H mass with neutral gas phase carbon
abundance (b) $\hii$ mass assuming $T=10^{4}$ K with elemental carbon
abundance $x_{C}\,=\,1.6\,\times\,10^{-4}$.  (c) $\h2$ mass with
neutral gas phase carbon abundance $x'_{C}\,=\,0.65\,\times\,10^{-4}$.
(d) Total gas mass traced by $\cii$ emission, including helium
contribution of $35\%$ by mass.
\label{masses}
\end{table}

The results are shown in Table\,\ref{masses}. The mass for the central
CND was calculated by taking all $\cii$ flux from the central pixel;
it might in actual fact contain a significant contribution from the
outflow. The outflow has an ionized hydrogen mass fraction of about
$16\%$. In the CND, that fraction is ten times lower but the actual
mass of gas associated with ionized carbon is eight times larger in
the CND than in the outflow.

\section{Discussion} 

\subsection{The extent and pressure of gas in the center}

The far-infrared fine-structure line maps in
Fig.\,\ref{linemaps} reveal a high-contrast compact core surrounded by
lower surface brightness emission from the resolved CND and the much
larger ETD.  Mid-infrared fine-structure lines, mostly from
species with a higher ionization potential, have been observed with
{\it ISO} and {\it Spitzer} (Sturm \etal 2002, Weedman \etal 2005,
Quillen $\etal$ 2008, and Ogle \etal 2010). They did not resolve the
CND but supported the results summarized in Table\,\ref{lineratios}.
The $\nev14\mu$m/24$\mu$m and $\siii18\mu$m/$33\mu$m ratios both
indicate electron densities less than a few hundred $\cc$. The neon
lines imply an average ionisation parameter log $U\,\approx\,-3.2$
(Sturm \etal. 2000, Dudik \etal 2007, Nagao \etal 2011), and a total
hydrogen column density $N_{H}>10^{21}\,\cm2$ (Pereira-Santaella \etal
2010). Assuming an electron temperature of about $10^{4}$ K,
appropriate to a fully ionized gas, the ionized gas pressure is
$n_{e} k T_{e}\approx10^{-12}$ N m$^{-2}$.

The compactness and structure of the bright core can at present only
be determined from high-resolution observations of near-infrared
lines. Ground-based Pa-$\beta$ and $\feii$ images showed ridges of
hydrogen emission extending from a compact core along the northern
radio jet over at least $4"$ (Krajnovi\'c $\etal$ 2007), and the
$\neii$ emission observed by Sturm $\etal$ (2002) originated in nuclear
region no more than $3"$ (55 pc) across (Siebenmorgen \etal 2004).
$K$-band $\h2$ observations (Israel $\etal$ 1990, and Israel 1998, his
Fig. 10) likewise suggested a size of $3"-4"$, as well as an enclosed
dynamical mass (stars+black hole) of $M(dyn)\geq5\times10^{8}$
M$_{\odot}$.  Detailed {\it HST} (Pa-$\alpha$, $\feii$; Schreier \etal
1998) and VLT-Sinfoni images ($\sivi$, Br-$\gamma$, $\feii$; Neumayer
$\etal$ 2007) revealed an unresolved ($d<0.2"$, <3.7 pc) peak at the
nucleus, surrounded by a small ($2.5''\times1.5"$, 45x30 pc) emission
nebula elongated along the jet axis.  The distribution of the excited
molecular hydrogen mapped by Neumayer $\etal$ (2007), also extends
along the radio jet. Its extent is similar to that of the ionic lines
but it is less concentrated towards the center.

Thus, the currently available high-resolution information shows that
the bright ionized gas at the center of the CND is contained within a
radius of at most 30 pc from the nucleus with the exception of more
extended emission along the radio jet that we ascribe to outflow (see
below). It is very likely that the emission in the far-infrared
fine-structure lines discussed in this paper follows the same pattern.
This extent is very close to that proposed for the CND central cavity
(Israel $\etal$ 1990), and we conclude that the ionized gas is limited
to the cavity, either filling it or tracing its edges. The neutral gas
outside the cavity has a much higher density and a much lower
temperature (Table\,\ref{lineratios}), implying again a pressure
$n_{o} k T_{k}\approx10^{-12}$ N m$^{-2}$. We thus find that in the
inner CND, the neutral gas just outside the cavity is in approximate
pressure equilibrium with the ionized gas inside the cavity.

\subsection{The outflow of gas from the center}

The PACS maps (Fig.\,\ref{linemaps}) clearly show the presence of
ionized gas outflow from the nucleus perpendicular to the CND plane
along the radio jet axis. In the nitrogen, lines this outflow
outshines even the central core, and it is also prominent in the
high-ionization $\oiv$ and $\nev$ lines mapped by Quillen $\etal$
(2008). Ionized emission traces the outflow over a projected distance
of about 220 pc, in the same position angle as the elongation of the
compact ionized gas traced by the near-infrared emission lines on
scales of a few tens of parsecs.  Thus, the structure seen in the
near-infrared lines represents an ionization cone rather than a
compact disk rotating perpendicularly to the CND as has sometimes been
proposed (cf. Schreier $\etal$ (1998).

Both the mid-infrared $\oiv$ and $\nev$ {\it Spitzer} images and the
far-infrared $\cii$, $\nii$ and $\niii$ PACS images also show an
apparent outflow feature towards the southwest. The axis of this
complementary feature is offset from that of the northeastern outflow,
and it is much less clearly defined. Its reality is, however, not in
doubt as it is seen in all the images. The difference between the two
sides of the ionized gas outflow resembles the larger-scale difference
between the synchrotron and X-ray emission from the Cen~A northern and
southern inner lobes. In both cases, we see a narrowly collimated
feature to the northeast and a much more poorly constrained feature to the
southwest.

As shown above, the outflow is also identified in the neutral gas
traced by $\ci$ atoms and excited CO molecules
(Figs.\,\ref{carbonmapfig} and \ref{pvplots}). The neutral gas seems
to be symmetrically distributed over the two sides of the outflow,
although it is really at the limit of our resolution.

\subsection{Shocked molecular gas in the center}

The dense gas in the compact central core, so close to a supermassive
black hole, is in a highly energetic state as shown by the
kinematics mapped at high resolution through near-infrared lines
(Neumayer $\etal$ 2007).

The $K$-band (1-0)$S(J)$ $\h2$ line ratios imply the presence either
of shock-excited gas at a temperature of $\leq$1000 K or of
fluorescent high-density gas irradiated by a strong UV radiation
fields (Israel $\etal$ 1990). However, the central region has
negligible star formation (Paper I, Radomski $\etal$ 2008). Shock
excitation of the gas is supported by the presence of shock indicators
such as ionized iron and silicon, and high $\feii$/Paschen line ratios
(cf. Marconi $\etal$ 2000, Krajnovi\'c $\etal$ 2007). The sharp rise
of the $\feii$/$\pii$ ratio with apertures up to $1.25"$ (Krajnovi\'c
$\etal$ 2007) suggests an increasingly dominating shock contribution
(See Terao $\etal$, 2016). In addition, the high surface brightness of
the 1-0 $S(1)$ (Israel $\etal$ 1990), $\feii$26$\mu$m and
$\siii$35$\mu$m lines (Sturm \etal 2002, Ogle \etal 2010) is likewise
characteristic for shock excitation and incompatible with the PDR
models of Kaufman \etal (2006) for any reasonable combination of
density n$_{\circ}$ and radiation field $G$. At least the $\siii$ flux
originates in a region less than $11"$ across, excluding emission from
either the outer CND or the outflow.

Ogle \etal (2010) found $\h2$ 0-0$S(J)$ fluxes an order of magnitude
stronger than the $K$-band (1-0)$S(J)$ flux. This suggests that most
of the warm molecular gas is excited by modest shock velocities well
below 20 $\kms$ at temperatures below 450 K (Draine \etal 1983;
Burton \etal 1992). Ogle \etal modeled their observed $\h2$
(0-0)$S(J)$ fluxes with three temperature components between 100 and
1500 K, assuming a source size of $3.7"\times3.7"$ equal to the {\it
  Spitzer} spectroscopic slit width. The actual source size is not
established by the observations and the modeled mass is critically
dependent on the assumed temperature of the `coolest' warm gas
component. When corrected to our assumed distance, their shocked
molecular hydrogen mass is $\sim11\%$ of all the molecular gas in the
whole CND (Paper I; This Paper, next section).  However, Fig. 10 in
Ogle $\etal$ showed that the measured $\h2$ line fluxes are consistent
with a two-temperature model at 290 K and 1500 K, lacking the
presumptive 100 K component (see also Quillen $\etal$ 2008).  This
reduced the mass of shocked molecular gas to $M(\h2)=0.4\times10^{6}$
M$_{\odot}$, corresponding to a much more reasonable fraction
($0.4\%$) of the total CND gas mass.

\subsection{Central mass budget }

How does the mass of gas associated with the $\cii$ emission relate to
that associated with neutral atomic carbon and molecular carbon
monoxide?  The lower angular resolution of the $\ci$ and CO
observations discussed in Sect. 3.2 does not allow us to determine the
mass of individual components in the same detailed way as we did for
$\cii$. Thus, we can only compare the total $\cii$ related CND gas
mass, $5.0\times10^{7}$ M$_{\odot}$, to the masses associated with CO
and $\ci$ from Table\,\ref{cicotab}, $(1.3-2.9)\times10^{7}$
M$_{\odot}$ and $(1.4-2.8)\times10^{7}$ M$_{\odot}$, respectively. We
adopt a CND total mass of dense gas
$M_{gas}(dense)=(9.1\pm0.9)\times10^{7}$ M$_{\odot}$.
M$_{\odot}$. This value is about $10\%$ higher but more accurate than
and consistent with the result from Paper I. It is about half of the
mass derived by Kamenetzky $\etal$ (2014) but, as they suggest, that
difference is easily explained by the contribution from the ETD in
their larger aperture.

Depending on assumptions, the fraction of CND gas directly associated
with CO is between $14\%$ (nominal case) and $17-30\%$ (reduced
cases); it is a minor fraction under any circumstance. The mass
associated with $\ci$ is similar, between $30\%$ in the nominal case
and $18-22\%$ in the reduced cases. Obviously, most of the mass,
between $55\%$ nominal and $50-65\%$ reduced, is associated with
ionized carbon rather than neutral carbon in any form. Thus, $\cii$ is
the best tracer of dense gas in the center of Cen~A, outperforming CO
by up to a factor four.

\subsection{Outflow dynamics}

In Section 4.2, we found a $\cii$-associated total gas mass for the
outflow in PACS pixel (2,1) of at least $3.4\times10^{6}$ M$_{\odot}$
and nominally $8.0\times10^{6}$ M$_{\odot}$. An even higher mass would
be found if either the temperature or the density is significantly
lower than we assumed.  As the assumed density is close to the
critical density, an increase has little effect: if it were a factor
of four higher the mass would still be about two thirds of what we
determined. However, if the density is a factor of four lower (\ie
log(n)=3.0) the implied mass would be more than a factor of two
higher. Furthermore, only $\sim60\%$ of the total mass in the CND is
associated with $\cii$. If that also holds for the outflow, its total
gas mass will be significantly higher. We therefore adopt a best value
$M_{out}(\cii)=1.2 (-0.6,+1.5)\times10^{7}$ M$_{\odot}$, a mass
uncertain by a factor two.

In the position-velocity maps of the submillimeter lines presented in
Section 3.2, we noted features that we interpreted as part of the
outflow better seen in the ionized carbon and nitrogen lines. In that
case, the northern outflow would have a projected radial velocity
$v'_{r}=60\pm10\,\kms$. The pixel size of $9.4"$ corresponds to a
projected linear size of $l'\,=\,174$ pc. If $i$ is the angle of the
outflow to the line of sight, the actual length is $l=l'/sin(i)$, the
space velocity is $v=v'_{r}/cos(i)$, and it takes a time
$t=l/v=(l'/v')tan(i)^{-1}$ for matter to clear the outflow pixel.
Unfortunately, the viewing angle $i$ is poorly known. If we assume
that the CND, whatever its true nature, is a spherically symmetrical
structure, its axial ratio of 0.5 suggests $i\approx60^{\circ}$.
Tingay $\etal$ (1998) deduced from the radio jet properties that $i$
should be $50^{\circ}$ or higher (up to $80^{\circ}$), Hardcastle
$\etal$ (2003) argued for a value less than $50^{\circ}$, preferably
about $15^{\circ}$, and M\"uller \etal (2014) found $i$ to lie between
$12^{\circ}$ and $45^{\circ}$. Fits of the velocity field of (ionized)
gas in the inner $1"-2"$ suggest outflow viewing angles
$i=25^{\circ}-45^{\circ}$ (Neumayer 2010).

If we take $i=50^{\circ}$, the outflow velocity is $95\pm15\,\kms$,
and the clearing time is $t_{50}$ = 2.5 million years.  If instead we
take $i=15^{\circ}$, the velocity is $60\pm10\,\kms$ but the clearing
time increases to $t_{15}$ = 11 million years. Thus, the uncertainty
in viewing angle $i$ is a major source of error. We adopt a viewing
angle $i=24^{\circ}(-9^{\circ},+26^{\circ})$, which corresponds to an
intermediate value for the clearing time $t_{24}$ = $6.5\pm3.1$
million years, again an overall uncertainty by a factor of two.

With these numbers, we deduced a mass outflow rate d$M$/d$t$ = 1.8
M$_{\odot}$/yr uncertain by a factor of three.  The (mostly atomic)
outflow in Cen~A is therefore modest compared to the molecular
outflows attributed to very luminous galaxies such as Mkn~231. The
situation observed in Cen~A is superficially similar to that seen in
500 pc surrounding the AGN of M~51 (less energetic by two to three
orders of magnitude), where Querejeta $\etal$ (2016) found that the
outflow also roughly balances the inflow with about one solar mass per
year each. They noted that the ionization cone is lacking in molecular
gas although such gas is found along its edges. However, the jet in
M~51 is only modestly inclined to the central disk
($i\approx15^{\circ}$) and their interaction appears to be more
complex than that in Cen~A with jet and outflow perpendicular to the
CND.  We note that the Cen~A outflow mass found by us is very similar
to the reduced mass, $M\,=\,2\times10^{6}$ M$_{\odot}$, of optically
thin molecular gas in the fast ($650\,\kms$) IC~5063 outflow derived
by Dasyra $\etal$ (2016).  The Cen~A outflow is also comparable to the
ionized gas outflows in other nearby AGN and Seyfert galaxies, as
summarized by Storchi-Bergmann (2014). These likewise extend over a
few hundred of parsecs, with total ionized gas masses very similar to
what we found, and mass outflow rates of 1–10 M$_{\odot}$.
Storchi-Bergmann (2014) noted that the mass outflow rates are 100 –
1000 times the AGN accretion rate, and suggested that the outflows are
due to mass loading of AGN jets as they move through the circumnuclear
ISM of the host galaxy. In Cen~A, the mass outflow rate is also more
than three orders of magnitude higher than the accretion rate of the
central black hole (d$M$/d$t$ = $1.2 (-0.3,+0.9)\times10^{-3}$
M$_{\odot}$/yr, Rafferty $\etal$ 2006), though mass loading of the AGN
jet does not appear to be the case.  The mass outflow is much more
efficient than black hole infall in emptying the circumnuclear cavity,
but the mechanism is not clear, nor is its relation to nuclear
feedback. At the present mass outflow rate, the entire CND would be
depleted of gas in a time somewhere between 15 and 120 million
years. Since the ETD is the remnant of a merger that most likely took
place $240\pm80$ million years ago (Quillen $\etal$ 1993; Struve
$\etal$ 2010), the CND depletion time is a significant fraction of the
time elapsed since the creation of the ETD; it may be comparable to
it.

Finally, we point out that only for viewing angles well above
$i=80^{\circ}$, the outflow space velocity would be high enough to
allow material to escape from the inner parts of Cen~A. For the more
probable lesser viewing angles, the outflowing gas is expected to slow
down, become turbulent and spread out over the inner kiloparsec or
so. What fraction, if any, will fall back to refuel the CND is at
present unclear.

\section{Conclusion}

We have used the Herschel PACS instrument and the ground-based APEX
telescope to map the inner $0.75\times0.75$ kpc of the giant
elliptical galaxy NGC 5128, hosting the FR-1 radio source, in various
lines of atomic carbon, oxygen, and nitrogen, and the J=4-3 transition
of molecular carbon monoxide. This has allowed us to deduce the
physical parameters of the interstellar medium in the central region
of a galaxy dominated by a supermassive black hole.  Although the
spatial resolutions $9.5"$-$13.5"$ provided are still unsurpassed at
these wavelengths, they only just allowed us to resolve the most
important structures in the Cen~A center.

The most conspicuous is the bright central object, about $20"$ in
diameter, that is called the circumnuclear disk (CND), although its
actual morphology and structure are unknown.  The continuum
emission from warm dust in the CND dominates the emission from the
nuclear point source only between wavelengths of 25$\mu$m and
300$\mu$m. Under the assumption of a single dust emission curve, the CND
temperature is 31 K. Uncertain by factors of two, its mass is about
$3.5\times10^{5}$ M$_{\odot}$ and its dust to gas ratio is
260. Neutral gas dominates the CND, only 10$\%$ is ionized. The
neutral gas has a three-quarter solar metallicity, a gas density of
about 20 000 $\cc$, and a temperature of 45 K, well above the
temperature of the CND dust. 

The CND does not appear to be heated by UV photons from newly formed
stars, because there is no sign of the intense star formation that is
seen in the extended thin disk (ETD) at greater distances to the
nucleus. The cooling of the CND is dominated by the $\cii$158$\mu$m
kine except in the center where the $\oi$63$\mu$m emission is strongest,
even though it suffers a great deal of absorption.

The CND has a total gas mass (including helium) of
$M_{gas}\,=\,9.1\times10^{7}$ M$_{\odot}$, with $10\%$ error.  Both
dense ($n\,=\,10^{4}-10^{5}$ $\cc$) and tenuous (density of a few
hundred) molecular gas occurs in the CND.  In the dense gas,
relatively little carbon is found in CO. More than half of the carbon
is ionized and emits in the $\cii$ line. We found nominal column density ratios
4:2:1 for $\cii$, $\ci$, and CO respectively.  Although only a few
resolution elements cover the CND, the ionized emission appears to be
more intense in the center than in the outer parts of the CND.
Further comparison with previously published observations of ionized
emission at mid-infrared and near-infrared wavelengths suggests that
much of the total ionized luminosity of the CND originates in a small
central cavity, only a few dozen parsec across, with edges coated by
shock-excited $\h2$. The shocked molecular gas mass is probably less
than a per cent of the total CND gas mass, but shocks and turbulence
nevertheless appear to dominate the central energetics.

An important result from the study described in this paper is the
discovery of an outflow of gas from the center of Cen~A along an
axis close to that of the northern X-ray/radio jet. The outflow was
just detected in the neutral gas traced by CO and $\ci$ emission that
suggest a projected outflow velocity of about 60 $\kms$, very much less
than the sub-relativistic speeds assigned to the radio jet. The
outflow was easily seen in the $\cii$, $\nii$, and $\niii$ emission
lines. Although the $\cii$ intensity peaks at the central position,
and gradually decreases along the outflow, the $\nii$ and $\niii$
intensities are highest in the outflow. There may be a less distinct
southern counterpart to the well-collimated northern outflow. Because
of the marginal resolution, and the tilted appearance of the CND, the
mass of gas in the outflow is difficult to establish. The outflow
projected clear of the CND has a mass of 8 million solar masses with
$30\%$ error.  The part of the outflow projected against the CND may
have a similar mass (which would reduce the CND mass quoted before to
by the same amount).  

The mass outflow rate is estimated to be about two solar masses per
year, uncertain by a factor of three. The uncertainty is dominated, in
equal parts, by possible errors in the mass calculation, and the
poorly known viewing angle that determines the projection effects.
Notwithstanding the uncertainty, the outflow rate is typically three
orders of magnitude higher than the estimated accretion rate of the
supermassive black hole. At the same time, it is close
to the mass outflow rates identified in nearby Seyfert galaxies.
Unless viewing angle are more extreme than seems likely at
present, the outflow material will not escape from NGC 5128
altogether, although most of it may not directly fall back onto the
CND.

\begin{acknowledgements}
  We gratefully acknowledge assistance provided in the early stages of
  this work by Edo Loenen, who supplied most of Table 4, as well as
  Figures 1 and 2. We also thank Renske Smit and Sarka Wykes for
  useful comments that led to improvements in the paper.
\end{acknowledgements}

\end{document}